\documentclass[%
 reprint,
 nofootinbib,
 amsmath,amssymb,
 aps,
]{revtex4-2}

\usepackage{graphicx}
\usepackage{dcolumn}
\usepackage{bm}
\usepackage{chemformula}
\usepackage{mathtools}

\usepackage[dvipsnames]{xcolor}

\definecolor{BerlinU9}{HTML}{F18800}

\usepackage[colorlinks=true]{hyperref}
\hypersetup{
    colorlinks = true,
    linkcolor = BerlinU9,
    urlcolor = BerlinU9,
    linkbordercolor = {white},
    citecolor = BerlinU9,
    }

\usepackage{orcidlink}
\begin{document}

\title{Numerical simulations of black hole-neutron star mergers \\ with equal and near-equal mass ratios}

\author{Ivan \surname{Markin}\,\orcidlink{0000-0001-5731-1633}$^{1}$}
\email{ivan.markin@uni-potsdam.de}
\author{Mattia \surname{Bulla}\,\orcidlink{0000-0002-8255-5127}$^{2,3,4}$}
\author{Tim \surname{Dietrich}\,\orcidlink{0000-0003-2374-307X}$^{1,5}$}

\affiliation{${}^1$ University of Potsdam, Institute of Physics and Astronomy, Haus 28, Karl-Liebknecht-Strasse 24/25, 14476, Potsdam, Germany}
\affiliation{${}^2$ Department of Physics and Earth Science, University of Ferrara, via Saragat 1, I-44122 Ferrara, Italy}
\affiliation{${}^3$ INFN, Sezione di Ferrara, via Saragat 1, I-44122 Ferrara, Italy}
\affiliation{${}^4$ INAF, Osservatorio Astronomico d’Abruzzo, via Mentore Maggini snc, 64100 Teramo, Italy}
\affiliation{${}^5$ Max Planck Institute for Gravitational Physics (Albert Einstein Institute), Am M\"uhlenberg 1, Potsdam 14476, Germany}

\date{\today}

\begin{abstract}
The detection of GW230529\_181500 suggested the existence of more symmetric black hole-neutron star mergers where the black hole mass can be as low as 2.6 times that of the neutron star. Black hole-neutron star binaries with even more symmetric mass ratios are expected to leave behind massive disks capable of driving bright electromagnetic transients like kilonovae.
Currently, there is only a limited number of numerical-relativity simulations of black hole-neutron star mergers in this regime, which are vital for accurate gravitational waveform models and analytical fitting formulas for the remnant properties. Insufficient accuracy of these may lead to misclassification of real events and potentially missed opportunities to locate their electromagnetic counterparts.
To fill this gap in the parameter space coverage, we perform simulations of black hole-neutron star mergers with mass ratios $q \in \{1, 1/2, 1/3\}$. We find the gravitational waveform models do not show good agreement with the numerical waveforms, with dephasing at the level of around $1$~rad at the merger. We find that the masses of the dynamical ejecta and disk are in good agreement with the available fitting formulas. The analytical formulas for the remnant black hole are in excellent agreement for the black hole mass, but are less accurate with the predictions for its spin. Moreover, we analyze the remnant disk structure and dynamics, deriving the rotation law and identifying global trapped $g$-mode density oscillations. We distinguish three types of accretion in the postmerger and find modulation of the accretion rate by the global oscillations of the disk.
Finally, we model the kilonova emission these systems would produce and find that most of them are potentially detectable by Vera C. Rubin Observatory within four days after merger, and by DECam within two days after merger if located at a distance of 200~Mpc.
\end{abstract}

\maketitle

\section{Introduction}
Multiple gravitational-wave (GW) signals from black hole-neutron star (BHNS) mergers have been detected to date: GW200105\_162426, GW200115\_042309~\cite{LIGOScientific:2021qlt}, and GW230529\_181500 (hereafter GW230529)~\cite{LIGOScientific:2024elc}.
According to stellar evolution models, black holes (BHs) are expected to have a mass of at least $5~M_\odot$, marking the so-called lower mass gap between the heaviest neutron stars (NSs) and the lightest BHs~\cite{Ozel:2010su,Farr:2010tu}. Therefore, most BHNS binaries formed from astrophysical BHs and NSs of typical mass of $1.4 \mathrm{M_\odot}$~\cite{Kiziltan:2013oja} (maximum mass of $ \sim 2.3 \mathrm{M_\odot}$~\cite{Romani:2022jhd}) are expected to have the mass ratio at most $q \sim 1/3.5$ ($1/2.2$), where $q= m_2/m_1$, and $m_1$,$m_2$ are the masses of the primary and secondary objects, respectively.
As GW230529 is inferred to most likely be a BHNS merger of $3.6~\mathrm{M_\odot}$ BH with a $1.4~\mathrm{M_\odot}$ NS ($q \approx1/2.6$)~\cite{LIGOScientific:2024elc}, this hypothesized lower mass gap was found to be populated with BHs.

The existence of BHNS systems with comparable component masses may indicate their formation through dynamical interactions in dense star clusters, while the effects of mass segregation there may lead to the production of BHNS systems where both the BH and NS have comparable masses~\cite{OLeary:2005vqo,Clausen:2012zu,Ye:2019xvf,Sedda:2020wzl}. BHNS systems with mass ratios $q \sim 1/2$ may also be second-generation systems, i.e., formed as a result of dynamical interaction of an NS with a remnant of an earlier binary neutron star (BNS) merger, whose mass is roughly double that of the initial NS in the binary~\cite{Gupta:2019nwj,Samsing:2020qqd,Mahapatra:2025agb}. On the other hand, symmetric BHNS systems with larger component masses may form from a binary NS system, where one of the NSs collapses, e.g., due to accretion and, thus, exceeding the maximum mass allowed by their equation of state (EOS)~\cite{Siegel:2022gwc}. Alternatively, such BHs might be an aftermath of a premature collapse of an NS~\cite{Abramowicz:2017zbp,East:2019dxt} or white dwarf triggered by a smaller BH, such as a primordial one, if those exist~\cite{Capela:2013yf,Takhistov:2017bpt,Genolini:2020ejw,Sasaki:2021iuc,Genolini:2020ejw,Richards:2021upu,Baumgarte:2024iby,Baumgarte:2024ouj}.
As recently estimated by Ref.~\cite{Baumgarte:2025syh}, a significant fraction of BHs in the NS-mass range can be produced by premature collapse of main-sequence stars triggered by capture of small primordial BHs.

\begin{table*}[tp]
    \centering
    \setlength{\tabcolsep}{4.5pt}
    \begin{tabular}{l|l|ccccc|cccc|ccc}

        &    &   &  &   &   &  &   &   &  &  &   & $M\Omega_{\mathrm{orb,ini}}$ & $e$  \\
        
        Name & CoRe  & $m_\mathrm{NS}$ & $m_\mathrm{BH}$& $q$ & $M$ & $\mathcal{M}_c$ & EOS & $\mathcal{C}_\mathrm{NS}$ &  $\Lambda_\mathrm{NS}$ & $\tilde{\Lambda}$ & $d$ & $\mathrm{[10^{-2}]}$ & $\mathrm{[10^{-4}]}$  \\

        \hline
\verb|DD2_mNS0.8_mBH1.6| & \verb|BAM:0227| & 0.802 & 1.6 & $1/2$ & 2.402 & 0.9745 & DD2 & 0.0908 & 12527 & 1594 & 30 & 2.046 & 7 \\
\verb|DD2_mNS1.2_mBH1.2| & \verb|BAM:0228| & 1.203 & 1.2 & $1$ & 2.403 & 1.0461 & DD2 & 0.1351 & 1590 & 799 & 31 & 1.957 & 4 \\
\verb|DD2_mNS1.2_mBH2.4| & \verb|BAM:0229| & 1.203 & 2.4 & $1/2$ & 3.603 & 1.4623 & DD2 & 0.1351 & 1590 & 203 & 45 & 2.046 & 7 \\
\verb|DD2_mNS1.4_mBH1.4| & \verb|BAM:0230| & 1.402 & 1.4 & $1$ & 2.802 & 1.2196 & DD2 & 0.1568 & 678 & 340 & 34 & 2.133 & 6 \\
\verb|DD2_mNS1.4_mBH2.8| & \verb|BAM:0231| & 1.402 & 2.8 & $1/2$ & 4.202 & 1.7047 & DD2 & 0.1568 & 678 & 86 & 50 & 2.191 & 3 \\
\verb|DD2_mNS2.2_mBH2.2| & \verb|BAM:0232| & 2.204 & 2.2 & $1$ & 4.404 & 1.9170 & DD2 & 0.2516 & 29 & 15 & 56 & 1.997 & 7 \\
\verb|DD2_mNS2.2_mBH4.4| & \verb|BAM:0233| & 2.204 & 4.4 & $1/2$ & 6.604 & 2.6795 & DD2 & 0.2516 & 29 & 4 & 81 & 2.101 & 1 \\
\verb|SLy_mNS1.2_mBH1.2| & \verb|BAM:0234| & 1.202 & 1.2 & $1$ & 2.402 & 1.0454 & SLy & 0.1547 & 805 & 403 & 29 & 2.147 & 5 \\
\verb|SLy_mNS1.2_mBH2.4| & \verb|BAM:0235| & 1.202 & 2.4 & $1/2$ & 3.602 & 1.4612 & SLy & 0.1547 & 805 & 102 & 43 & 2.180 & 4 \\
\verb|SLy_mNS1.4_mBH1.4| & \verb|BAM:0236| & 1.397 & 1.4 & $1$ & 2.797 & 1.2175 & SLy & 0.1800 & 311 & 155 & 34 & 2.127 & 5 \\
\verb|SLy_mNS1.4_mBH2.8| & \verb|BAM:0237| & 1.397 & 2.8 & $1/2$ & 4.197 & 1.7015 & SLy & 0.1800 & 311 & 39 & 50 & 2.187 & 4 \\
\verb|SLy_mNS1.4_mBH4.2| & \verb|BAM:0238| & 1.397 & 4.2 & $1/3$ & 5.597 & 2.0488 & SLy & 0.1800 & 311 & 14 & 66 & 2.217 & 2 \\
    \end{tabular}
    \caption{Binary parameters of the simulated configurations. First column: configuration name. Second column: waveform identifier in CoRe database~\cite{Dietrich:2018phi,Gonzalez:2022mgo}. First column group: Arnowitt–Deser–Misner (ADM) mass of the NS in isolation $m_\mathrm{NS}$, Christodolou mass of the BH $m_\mathrm{BH}$, mass ratio $q=m_\mathrm{NS}/m_\mathrm{BH}$, total mass of the binary $M=m_\mathrm{BH} + m_\mathrm{NS}$ at the infinite separation, and chirp mass $\mathcal{M}_c$. Second column group: name of the employed EOS, the NS compactness $\mathcal{C}_\mathrm{NS}$, quadrupolar tidal deformability of the NS $\Lambda_{\mathrm{NS}}$, effective tidal deformability of the binary $\tilde{\Lambda}$. Third column group: initial binary separation $d$, mass-scaled initial orbital frequency $M\Omega_{\mathrm{orb,ini}}$, and residual eccentricity in the initial data $e$.}
    \label{table:configurations}
\end{table*}

The GW230529 event has revived interest in simulating BHNS mergers with more symmetric mass ratios using numerical relativity~(NR)~\cite{Foucart:2019bxj,Martineau:2024zur,Matur:2024nwi}, as it enables detailed modeling of compact objects within the fully general-relativistic framework.
Moreover, the BHNS systems with symmetric masses, $q \approx 1$, tend to have more massive accretion disks~\cite{Foucart:2018rjc}, whereas for the cases with a nonspinning BH, the disk mass peaks for systems with $q \in [1/2, 1]$~\cite{Hayashi:2020zmn}. When the BH has a larger prograde spin, heavier disks are produced, with an extended distribution to more asymmetric mass ratios~\cite{Matur:2025avh}.
On the other hand, the mass of dynamical ejecta peaks at $q \sim 1/2$ for systems with nonspinning BHs, and extends to more asymmetric mass ratios when the BH has larger prograde spin~\cite{Kruger:2020gig}.
The combination of these two facts makes systems with mass ratios between $1/2$ and $1$ promising sources of electromagnetic counterparts, such as kilonovae and gamma-ray bursts~\cite{Narayan:1992iy,Li:1998bw,Janka:1999qu,Rosswog:2005su,Lee:2007js,Metzger:2019zeh}.

One motivation for investigating symmetric BHNS systems in NR is the verification of the accuracy of gravitational waveform models. 
The number of BHNS NR waveforms available for waveform model calibration is currently rather limited, with most of the NR waveforms having mass ratios $q \lesssim 1/3$~\cite{Thompson:2020nei,Matas:2020wab,Gonzalez:2022prs}, and only two publicly available NR waveforms for symmetric BHNS systems~\cite{Foucart:2019bxj}. Recently, the calibration range of some waveform models was extended to $q \sim 1/2$ systems~\cite{Gonzalez:2025xba}, although with shorter waveforms and with residual eccentricity.
A systematic study of low-mass BHNS mergers~\cite{Hayashi:2020zmn} covered systems with nonspinning components with a fixed NS mass and varying mass ratio $q \in [ 1/4.4, 1/1.5]$ and the EOS, though being focused on the disk and ejecta properties, no long waveforms were produced.
Besides, the region of the parameter space with equal mass ratio and various NS masses remains sparsely populated by NR simulations. 
Underrepresentation of these symmetric BHNS configurations in calibration sets of waveform models may reduce their accuracy in this regime, leading to biases and misinterpretation of the source properties of the real signals, if such systems exist. For example, a study of a BHNS merger with a subsolar BH~\cite{Markin:2023fxx} revealed that waveform models show high discrepancies with the NR data for more mass-symmetric systems. 

Motivated by these observational and theoretical developments, we perform NR simulations of nonspinning BHNS systems with mass ratios $q \in \{1, 1/2, 1/3\}$, two different equations of state (EOS), and various NS masses. We produce long and accurate gravitational waveforms, which we compare to the current BHNS waveform models. We verify the analytical fitting formulas for the remnant properties, such as BH mass and spin, mass of the dynamical ejecta, and the disk. In relevance to long-term modeling of the disk and potential electromagnetic counterparts, we analyze the morphology and dynamics of the remnant accretion disks formed, their rotation laws, global density oscillations, and the accretion behavior. Finally, we model the potential kilonova signals from these systems and their detectability by wide-field surveys.

The paper is structured as follows.
In Sec.~\ref{section:simulations}, we describe the selected configurations and the numerical methods.
In Sec.~\ref{section:gravitational_waveforms}, we discuss the gravitational waveforms obtained from our simulations and compare them with the BHNS waveform models.
In Sec.~\ref{section:remnant_properites}, we discuss properties of the remnant, such as ejecta, disk, and remnant BH.
In Sec.~\ref{section:disk_structure_accretion}, we study the remnant disk structure, its rotation laws, global density oscillations, and the accretion dynamics.
In Sec.~\ref{section:kilonova}, we model the kilonova emission associated with the mergers.
We conclude in Sec.~\ref{section:conclusion}.

Throughout the paper, we use geometric units for
all the physical quantities, i.e., $G=c=M_\odot=1$, unless
stated otherwise. Hereafter, we define merger time as the time when the GW strain amplitude reaches its maximum.

\section{Simulated configurations}
\label{section:simulations}
We simulate a set of BHNS configurations with $q \in \{1,1/2,1/3\}$, targeting NS masses $m_{\mathrm{NS}} \in \{0.8,1.2,1.4,2.2\}$, and zero spin for both components.
For the description of the NS matter, we select two representative equations of state (EOSs), softer SLy~\cite{Douchin:2001sv} and stiffer DD2~\cite{Hempel:2009mc,Typel:2009sy,Typel:2013rza}, and use a piecewise-polytropic fit for both (Ref.~\cite{Read:2008iy} for SLy and Refs.~\cite{Oertel:2016bki,CompOSECoreTeam:2022ddl} for DD2). To incorporate the thermal effects, we augment the cold piecewise-polytropic EOS with the ideal-gas extension~\cite{Bauswein:2010dn} with the adiabatic exponent $\Gamma_{\text{th}} = 1.75$.
We exclude the SLy configuration with $m_\mathrm{NS}=2.2$, as it exceeds the maximum supported mass for a nonspinning NS.
For $m_\mathrm{NS}=0.8$, we consider only one configuration with $q=1/2$ with the DD2 EOS to study the effect of very high tidal deformabilities. Simulating an equal-mass setup for this NS mass, i.e., with $m_\mathrm{BH}=0.8$, would be computationally too expensive as it would require additional refinement levels around the BH; cf.~\cite{Markin:2023fxx} for a case study of BHNS merger simulation with subsolar-mass BH.
We also consider only one configuration with $q=1/3$, so that it falls within the parameter range of simulations of such systems already studied in the literature.

We find that the evolved baryonic masses of the NSs differ from the target ones by $
\sim 10^{-3} \text{M}_\odot$ due to the interpolation of the initial data onto the finite-volume grid. We recalculate the NS parameters, such as gravitational mass and tidal deformability, based on the Tolman-Oppenheimer-Volkoff solutions for the interpolated baryonic mass from the evolution grid. Throughout the text, we refer to the target component masses for simplicity.

Due to a mismatch in the DD2 EOS prescriptions in the initial data and evolution, where the Stellar Collapse~\cite{OConnor:2009iuz} variant was used in the initial data instead of the intended one from CompOSE~\cite{CompOSECoreTeam:2022ddl}, these configurations start with NSs in a slightly perturbed state, and reequilibrate within the first $3$~ms. While the gravitational masses remain largely unchanged, the tidal deformabilities have a maximum change of $3$~\% from the intended ones. We find a negligible impact of the NS perturbations on the waveform, verified by rerunning the corrected simulations for a selected number of cases.

We evolve each configuration for around nine orbits. The configurations and their properties are listed in Tab.~\ref{table:configurations}.

\subsection{Initial data and evolution}
\label{section:initial_data_and_evolution}
We construct quasicircular initial data for the configurations using \textsc{FUKA}~\cite{Papenfort:2021hod}, which is based on the \textsc{Kadath} spectral library~\cite{Grandclement:2009ju}. \textsc{FUKA} solves the Einstein constraint equations using the extended conformal thin sandwich (XCTS) formulation~\cite{York:1998hy,Pfeiffer:2002iy}. On each configuration, we perform the eccentricity reduction procedure until the residual eccentricity falls below $10^{-3}$~\cite{Papenfort:2021hod,Pfeiffer:2007yz}. The residual eccentricities are listed in the Tab.~\ref{table:configurations}.

To dynamically evolve the systems in the general-relativistic hydrodynamics framework, we use \textsc{BAM}~\cite{Bruegmann:2006ulg,Thierfelder:2011yi,Dietrich:2015iva,Bernuzzi:2016pie,Dietrich:2018phi}. There, the spacetime is evolved using conformal Z4 (Z4c) reformulation~\cite{Hilditch:2012fp} of Einstein field equations, combined with the moving-puncture gauge~\cite{Campanelli:2005dd,Baker:2005vv}.
For the relativistic hydrodynamics, we use the Valencia formulation~\cite{Marti:1991wi,Banyuls:1997zz,Anton:2005gi,Font:2008fka}, employing high-resolution shock-capturing methods with the local Lax-Friedrichs (LLF) Riemann solver and using \mbox{WENO-Z} primitive variable reconstruction~\cite{2008JCoPh.227.3191B}. 

The computational grid in \textsc{BAM} forms a tree of $L$ nested Cartesian refinement levels. For each finer level, the resolution is two times higher, so for each level $l=\{0,1,...,L-1\}$, the grid spacing is $h_l = 2^{-l} h_0$. While the three outermost refinement levels are fixed, all finer levels are allowed to move. Each of the moving levels can contain multiple subgrids (boxes) that are constructed to fully cover each of the compact objects and set to follow their trajectories. Thus, the NS is always contained within the eighth level ($l=7$), while the BH is fully covered by additional levels so that the finest level provides the highest resolution for the BH. Once these boxes cover intersecting regions of the computational domain, they are merged into one, retaining the resolution. As the resolution of the simulation is primarily limited by that of the finest levels, we refer to the simulation resolution as $\mathrm{R}n$, where $n$ is the number of grid points along each dimension at the finest level. We perform all simulations at three distinct resolutions: R96, R144, and R192. We refer to the results of the highest-resolution (R192) runs as the main results throughout the paper and use the other resolutions for the convergence studies in Appendix~\ref{section:convergence}.
Further grid setup details are listed in ~Tab.~\ref{tab:numerical_setup} in the Appendix.
We perform time integration using a fourth-order Runge-Kutta scheme with Courant-Friedrichs-Lewy factor set to $0.25$, and employ the Berger-Colella scheme~\cite{Berger:1989a} for local time stepping, and maintain the conservation of the baryonic mass, energy, and momentum across the refinement levels~\cite{Berger:1989a,Dietrich:2015iva}.

\begin{figure*}[htp]
    \centering
    \includegraphics[width=1.0\linewidth]{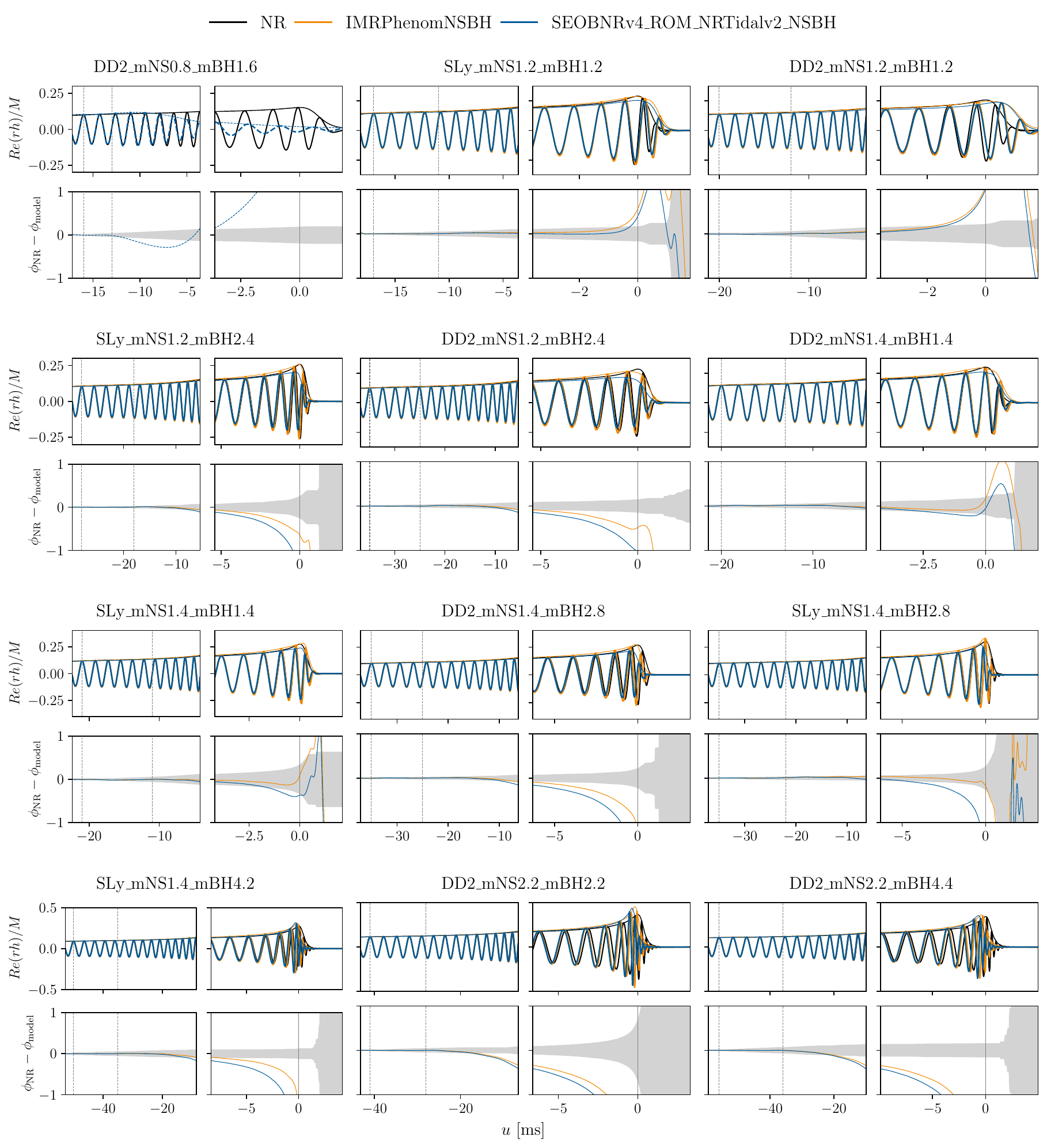}

    \caption{Numerical waveforms and their dephasing with the waveform models. {Upper panels:} real part of the (2,2)-mode of the GW strain. {Lower panels:} phase difference $\phi_\mathrm{NR} - \phi_\mathrm{model}$ between the numerical waveform and each of the waveform models. The gray area represents the NR waveform error, estimated as the absolute phase difference between the waveform phases at the highest and medium resolutions, plus the mass drift error caused by the initial data interpolation. The waveform model for the DD2\_mNS0.8\_mBH1.6 is shown in dashed lines to highlight that it was used outside of its allowed parameter space.}
    \label{figure:waveforms}
\end{figure*}

\section{Gravitational waveforms}
\label{section:gravitational_waveforms}
We extract the GWs from our NR simulations at multiple extraction spheres with different coordinate radii $r_{\text{extr}} \in \{600,700,800,900,1000\}$. To reduce the gauge effects in the waveforms, we perform second-order perturbative extrapolation of finite-radii waveforms using \textsc{scri}~\cite{Iozzo:2020jcu,boyle_2025_17080831}, factoring in the values of the average lapse at and areal radii of the extraction spheres. Additionally, we correct the waveforms to remove the mode-mixing effect caused by the center-of-mass drift~\cite{Woodford:2019tlo}, cf.~Appendix~\ref{section:com_correction}.

In Fig.~\ref{figure:waveforms}, we present all twelve waveforms at the highest resolution, R192, and compare them with BHNS waveform models\footnote{Because the aforementioned mass drift due to the initial data interpolation, the NS mass can be lower than that of the BH, which is not allowed by both waveform models. In these cases, we reset the NS mass to the mass of the BH.} \verb|IMRPhenomNSBH|~\cite{Thompson:2020nei} and \verb|SEOBNRv4_ROM_NRTidalv2_NSBH|~\cite{Matas:2020wab}.

In the case of \verb|DD2_mNS0.8_mBH1.6|, the tidal deformability of the NS is higher than the maximum allowed by both models ($5000$). To illustrate discrepancies caused by using a waveform model outside of its allowed parameter space, we remove the tidal deformability bound check in \verb|SEOBNRv4_ROM_NRTidalv2_NSBH|. We do not plot such a waveform for \verb|IMRPhenomNSBH|, as it becomes noisier than \verb|SEOBNRv4_ROM_NRTidalv2_NSBH| to be meaningful.
We align the waveform models with the numerical waveforms by minimizing the following integral over the alignment window $[t_0, t_1]$:

\begin{equation}
    \mathcal{I} = \int_{t_0}^{t_1} dt~|\phi_\mathrm{model}(t+\delta t) + \delta \phi - \phi_\mathrm{NR}(t)|^2,
\end{equation}
where $\phi_\mathrm{model}$ and $\phi_\mathrm{NR}$ are the waveform model and NR GW phases, and $\delta \phi$, $\delta t$ are the sought phase and time shifts.

In the vast majority of cases, both waveform models do not show good agreement with the NR data, and have $\mathcal{O}(1~\mathrm{rad})$ dephasing during late inspiral, which exceeds the NR error already at around three GW cycles before merger. 
The best agreement is achieved for \verb|DD2_mNS1.4_mBH1.4|, \verb|SLy_mNS1.2_mBH1.2|, and \verb|SLy_mNS1.4_mBH1.4|, where the phase error remains within the NR errors throughout most of the inspiral. The highest disagreement is produced for the \verb|DD2_mNS0.8_mBH1.6| waveform, where the phase error leaves the error band of the NR data immediately after the alignment window, and becomes larger than $1$~rad at two GW cycles before merger. The model predicts the merger around $10$~ms earlier than in the NR data, and has an extended ringdown stage. 
The dephasing between the models and NR data does not always have the same sign: equal-mass systems with the component mass of $1.2~\mathrm{M_\odot}$ show faster phase evolution in NR data, while for all others the opposite trend is true. This correlates with the fact that most models with faster phase evolution also predict earlier merger times.
Across all the cases, the \verb|IMRPhenomNSBH| waveforms consistently have faster phase evolution than the \verb|SEOBNRv4_ROM_NRTidalv2_NSBH| ones.

\section{Remnant properties}
\label{section:remnant_properites}
We list various remnant properties in Tab.~\ref{table:remnant_properties}, and discuss them separately below.

\begin{table}[!tp]
    \centering
    \begin{tabular}{l|r|r|cc}
        Name & $M_{\mathrm{ej}}^{\mathrm{dyn}}$ & $M_{\mathrm{disk}}$  & $M_{\mathrm{BH}}^{\mathrm{rem}}$  &  $\chi_{\mathrm{BH}}^{\mathrm{rem}}$ \\
        \hline
\verb|DD2_mNS0.8_mBH1.6| & $4.40 \times 10^{-2}$ & $1.17 \times 10^{-1}$ & $2.215$ & $0.61$ \\
\verb|SLy_mNS1.2_mBH1.2| & $< 1 \times 10^{-4}$ & $4.24 \times 10^{-2}$ & $2.323$ & $0.84$ \\
\verb|DD2_mNS1.2_mBH1.2| & $2.35 \times 10^{-4}$ & $9.52 \times 10^{-2}$ & $2.278$ & $0.85$ \\
\verb|SLy_mNS1.2_mBH2.4| & $1.85 \times 10^{-3}$ & $9.04 \times 10^{-2}$ & $3.447$ & $0.67$ \\
\verb|DD2_mNS1.2_mBH2.4| & $1.92 \times 10^{-2}$ & $1.27 \times 10^{-1}$ & $3.403$ & $0.65$ \\
\verb|DD2_mNS1.4_mBH1.4| & $< 1 \times 10^{-4}$ & $3.92 \times 10^{-2}$ & $2.716$ & $0.84$ \\
\verb|SLy_mNS1.4_mBH1.4| & $< 1 \times 10^{-4}$ & $2.92 \times 10^{-3}$ & $2.735$ & $0.81$ \\
\verb|DD2_mNS1.4_mBH2.8| & $3.29 \times 10^{-3}$ & $8.33 \times 10^{-2}$  & $4.034$ & $0.67$ \\
\verb|SLy_mNS1.4_mBH2.8| & $< 1 \times 10^{-4}$ & $1.32 \times 10^{-2}$ & $4.077$ & $0.67$ \\
\verb|SLy_mNS1.4_mBH4.2| & $< 1 \times 10^{-4}$ & $6.19 \times 10^{-4}$ & $5.453$ & $0.55$ \\
\verb|DD2_mNS2.2_mBH2.2| & $< 1 \times 10^{-4}$ & $< 1 \times 10^{-4}$ & $4.228$ & $0.72$ \\
\verb|DD2_mNS2.2_mBH4.4| & $< 1 \times 10^{-4}$ & $< 1 \times 10^{-4}$ & $6.352$ & $0.63$ \\
    \end{tabular}
    \caption{Dynamical ejecta mass $M_{\mathrm{ej}}^{\mathrm{dyn}}$, baryonic mass of the accretion disk $M_{\mathrm{disk}}$, remnant BH mass $M_{\mathrm{BH}}^{\mathrm{rem}}$ and spin $\chi_{\mathrm{BH}}^{\mathrm{rem}}$ for the binary configurations. The disk mass and the BH properties are calculated at $10$~ms after merger.}
    \label{table:remnant_properties}
\end{table}

\subsection{Dynamical ejecta}
\label{section:dynamical_ejecta}
\begin{figure}[!htp]
    \centering
    \includegraphics[width=1\linewidth]{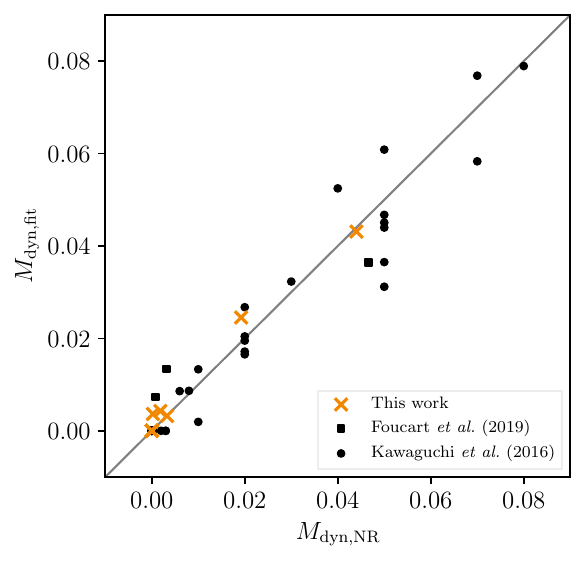}
    \caption{The predictions of the dynamical ejecta mass of the fitting formula of Ref.~\cite{Kruger:2020gig} plotted against the corresponding simulation data of this work and of Refs.~\cite{Kawaguchi:2015bwa,Foucart:2019bxj}.}
    \label{figure:ejecta_dyn_fit}
\end{figure}

Motivated to refine the analytical predictions of the merger parameters using our simulations, we examine the agreement with the fitting formula of Ref.~\cite{Kruger:2020gig} for the mass of the dynamical ejecta. To distinguish the ejecta from the bound matter, we use the geodesic criterion, $u_t < -1$, where $u_t$ is the covariant time component of the fluid four-velocity, together with a requirement of positive radial velocity. In Fig.~\ref{figure:ejecta_dyn_fit}, we compare the predictions of the fitting formula with the numerical data of our simulations. The predicted masses of the dynamical ejecta exhibit good agreement with the numerical data, with smaller errors than most of the configurations that are used for calibration of the fitting formula~\cite{Kawaguchi:2015bwa,Foucart:2019bxj}, also plotted in Fig.~\ref{figure:ejecta_dyn_fit}.
For most cases considered here, the dynamical ejecta mass is below $10^{-2}~\mathrm{M_\odot}$ and thus smaller than the error of the fitting formula itself. For the configurations with the dynamical ejecta mass above $10^{-3}~\mathrm{M_\odot}$, the numerical error in the ejecta mass (estimated at the difference between high and medium resolution values) is on average $3$\%, and always below $5$\%.

\subsection{Disk}
We calculate the disk mass as the baryonic mass of the bound matter at $10$~ms after merger, and list the values in Tab.~\ref{table:remnant_properties}.
In most cases, the disk mass is multiple orders of magnitude higher than the dynamical ejecta mass, except in several cases where they differ only by one order of magnitude. This fact makes the matter ejected by the subsequent disk winds driven by neutrino radiation and the magnetic field the most dominant contributor to the kilonova emission~\cite{Nakar:2019fza}. Here, however, we do not model these winds; yet, the disk properties remain useful for calibrating the analytical fitting formulas and the simplified kilonova modeling we perform in this work. Generally, the fitting formulas for disk mass are essential for estimating the kilonova brightness and its detectability~\cite{Hinderer:2018pei,Coughlin:2018fis,Raaijmakers:2021slr,Chase:2021ood,Kunnumkai:2024qmw}, as well as for waveform models to determine the waveform shutoff frequency, e.g., cf.~\cite{Thompson:2020nei}.

\begin{figure}[!tp]
    \centering
    \includegraphics[width=1\linewidth]{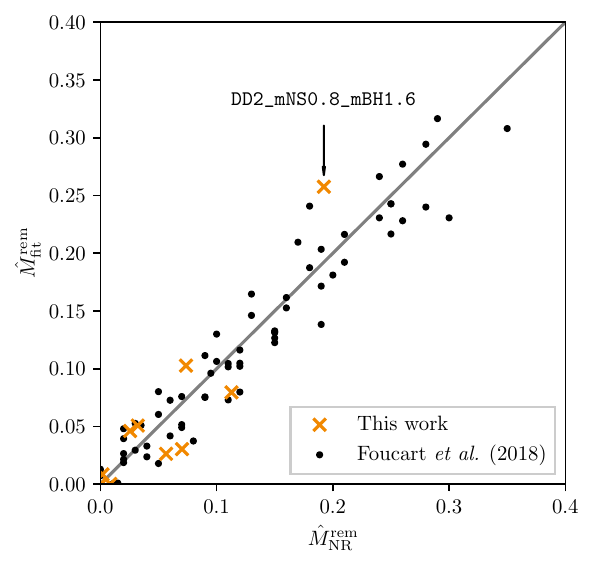}
    \caption{Remnant baryonic mass normalized by the initial baryonic mass of the NS predicted by the fitting formula of Ref.~\cite{Foucart:2018rjc}, $\hat{M}^\mathrm{rem}_\mathrm{fit}$, plotted against the values obtained from the simulations in this work, $\hat{M}^\mathrm{rem}_\mathrm{NR}$. For comparison, the configurations used for calibration of the Foucart~\textit{et al.} (2018)~\cite{Foucart:2018rjc} model are plotted additionally.}
    \label{figure:Mremnant}
\end{figure}

\begin{table*}[tp]
    \centering
    \begin{tabular}{l|ccc|ccc||c}
        Name & $M_{\mathrm{BH}}^{\mathrm{rem}}$ & $\Delta M_{\mathrm{BH}}^{\mathrm{rem}}$ \cite{Gonzalez:2022prs}  & $\Delta M_{\mathrm{BH}}^{\mathrm{rem}}$ \cite{Gonzalez:2025xba} & $\chi_{\mathrm{BH}}^{\mathrm{rem}}$   &  $\Delta \chi_{\mathrm{BH}}^{\mathrm{rem}}$ \cite{Gonzalez:2022prs} &  $\Delta \chi_{\mathrm{BH}}^{\mathrm{rem}}$  \cite{Gonzalez:2025xba} & $\kappa$,~[Hz] \\
        \hline
\verb|DD2_mNS0.8_mBH1.6| & 2.215 & 1.2\% & -0.7\% & 0.61 & -1.5\% & -45.2\% & 544 \\
\verb|SLy_mNS1.2_mBH1.2| & 2.323 & -2.1\% & -2.1\% & 0.84 & -5.3\% & -7.3\% & 719 \\
\verb|DD2_mNS1.2_mBH1.2| & 2.278 & -1.2\% & -2.7\% & 0.85 & -7.1\% & -14.2\% & 747 \\
\verb|SLy_mNS1.2_mBH2.4| & 3.447 & 0.1\% & 0.0\% & 0.67 & 0.2\% & -1.3\% & 376 \\
\verb|DD2_mNS1.2_mBH2.4| & 3.403 & 0.4\% & -0.6\% & 0.65 & 1.2\% & -2.7\% & 372 \\
\verb|DD2_mNS1.4_mBH1.4| & 2.716 & -2.0\% & -1.7\% & 0.84 & -5.5\% & -6.3\%  & 615 \\
\verb|SLy_mNS1.4_mBH1.4| & 2.735 & -1.7\% & -1.1\% & 0.81 & -4.7\% & -4.0\% & 580 \\
\verb|DD2_mNS1.4_mBH2.8| & 4.034 & 0.0\% & 0.2\% & 0.67 & 0.4\% & -0.4\% & 321\\
\verb|SLy_mNS1.4_mBH2.8| & 4.077 & -0.2\% & -0.0\% & 0.67 & -0.6\% & -0.5\%  & 318 \\
\verb|SLy_mNS1.4_mBH4.2| & 5.453 & 0.3\% & 0.2\% & 0.55 & 1.1\% & 0.6\% & 207 \\
\verb|DD2_mNS2.2_mBH2.2| & 4.228 & -0.1\% & -0.6\% & 0.72 & -2.7\% & -2.9\% & 328 \\
\verb|DD2_mNS2.2_mBH4.4| & 6.352 & 0.5\% & 0.1\% & 0.63 & 0.2\% & 0.0\% & 194 \\
             \end{tabular}
    \caption{Remnant mass $M_{\mathrm{BH}}^{\mathrm{rem}}$ and spin $\chi_{\mathrm{BH}}^{\mathrm{rem}}$ for our data with their corresponding relative differences $\Delta M_{\mathrm{BH}}^{\mathrm{rem}}$ and $\Delta \chi_{\mathrm{BH}}^{\mathrm{rem}}$ when compared to the values predicted by the analytical fitting formulas from Refs.~\cite{Gonzalez:2022prs,Gonzalez:2025xba}. The last column represents the epicyclic frequency $\kappa$ calculated using Eq.~\eqref{eq:epicyclic_frequency}.}
    \label{table:remnant_BH_fits}
\end{table*}

One of such widely used analytical formulas is that of Ref.~\cite{Foucart:2018rjc}. It, however, fits the masses of the entire baryonic remnant, which includes both disk and ejecta. As noted above, the contribution of the ejecta to the baryonic remnant mass is often small, and the formula can be used as a proxy for the disk mass in such cases.
In Fig.~\ref{figure:Mremnant}, we compare the predictions of the Foucart~\textit{et al.}~(2018)~\cite{Foucart:2018rjc} formula for the baryonic remnant mass with the values from our simulations here. All configurations show good agreement with the fit. The only marginal outlier is the system with a subsolar NS, \verb|DD2_mNS0.8_mBH1.6|, for which the mass of the baryonic remnant is overpredicted by the formula by around $34$\%. We suggest that this is most likely due to the NS's low compactness, which lies well outside the calibration region of the fit.
The numerical error of the baryonic mass of the remnant, estimated at the difference between the high and medium resolutions, is always below $10^{-3}~\mathrm{M_\odot}$ and remains within $4\%$ for all configurations except \verb|SLy_mNS1.4_mBH2.8|, where it reaches $11\%$. This error is significantly smaller than the error of the fitting formula.

\subsection{Black hole}
We calculate the mass $M_{\mathrm{BH}}^{\mathrm{rem}}$ and dimensionless spin parameter $\chi_{\mathrm{BH}}^{\mathrm{rem}}$ of the remnant BH as the mass of the apparent horizon at $10$~ms after the merger, and present the values in Tab.~\ref{table:remnant_properties}.

The highly spinning ($\chi > 0.7$) remnant BHs are produced by the equal-mass ratio configurations due to the reduced radiation of angular momentum for unequal mass systems~\cite{Buonanno:2007sv,Lousto:2009mf,Varma:2019csw}.

We compare our NR values for the BH remnant properties with the fitting formulas of Refs.~\cite{Gonzalez:2022prs,Gonzalez:2025xba} for the BH remnant properties of BHNS mergers, and present the results in Tab.~\ref{table:remnant_BH_fits}. As we noted in our recent BHNS merger study~\cite{Markin:2025oeo}, the updated formula from Ref.~\cite{Gonzalez:2025xba} provides better prediction for the BH mass but not for the spin. Here, the updated formula provides a more accurate estimate for the BH mass than its predecessor. The only cases where the error worsened by a factor of roughly two are \verb|DD2_mNS1.2_mBH1.2| and \verb|DD2_mNS1.2_mBH2.4|.
For the BH spin, the updated formula disagrees more with the NR data in most cases, especially those with larger tidal deformability, i.e., with the DD2 EOS. Notably, for the subsolar NS case, \verb|DD2_mNS0.8_mBH1.6|, the predicted spin disagrees with the numerical data by almost 50\%.

We highlight that these fitting formulas are essential components of the waveform models, and large disagreements lead to a reduction of their accuracy.

\begin{figure*}[!htp]
    \centering
    \includegraphics[width=1\linewidth]{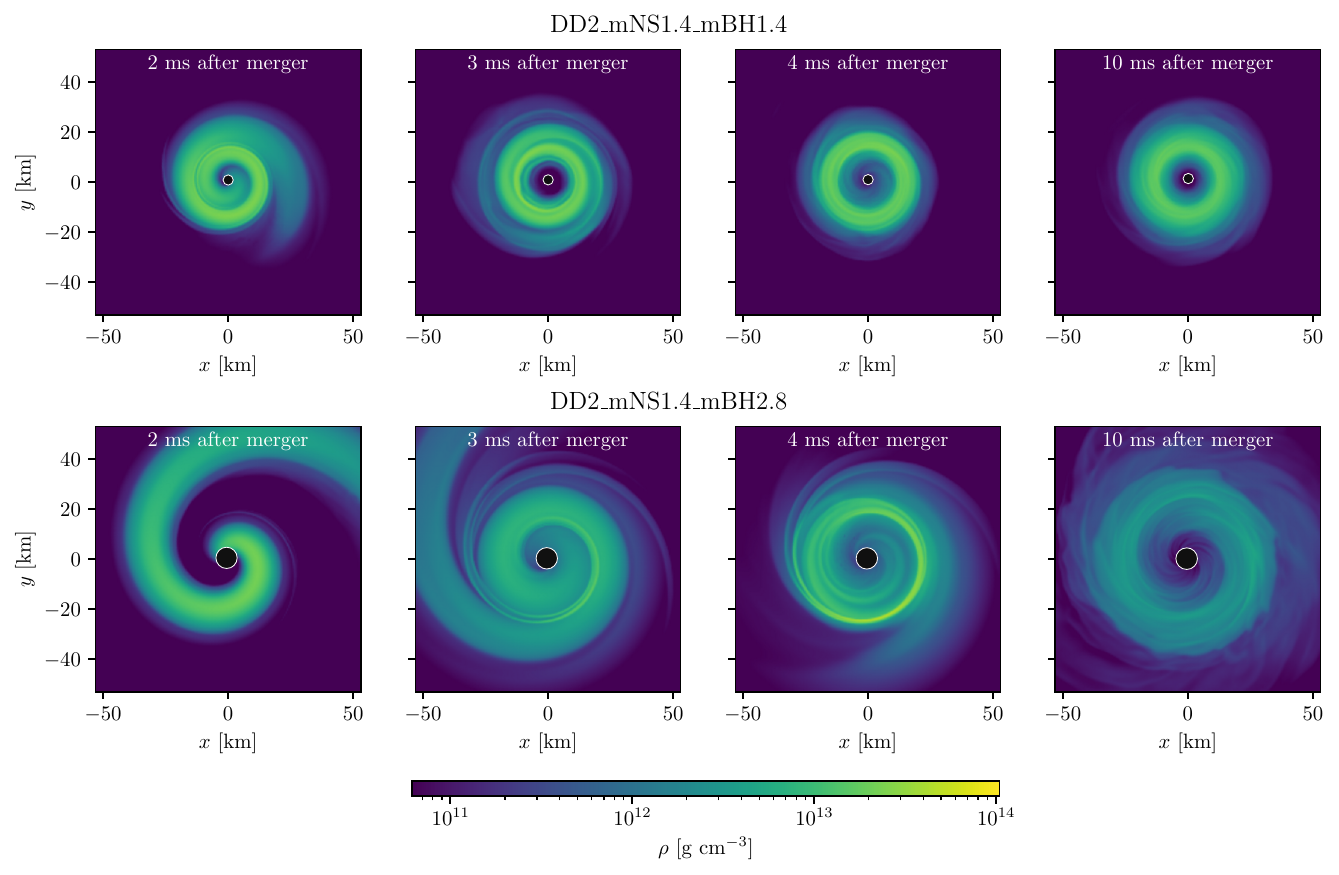}
    \caption{Disk matter density in the equatorial plane at selected times for two BHNS configurations with $m_\mathrm{NS}=1.4$ and $q=\{1, 1/2\}$, with the black circles with white outline showing the apparent horizons. In the equal-mass ratio, the disk circularizes almost immediately, whereas in the $q=1/2$ case, the disk is highly perturbed by spiral density wave interaction and the fallback matter.}
    \label{fig:disk_over_time}
\end{figure*}

\section{Disk structure and accretion}
\label{section:disk_structure_accretion}

The BHNS remnant disks are of particular interest for simulation at longer timescales, as additional matter will become unbound later on and will become the main driver of the kilonova emission. Continuing NR simulations for these timescales can be prohibitively expensive; therefore, to further evolve the disk, one must simplify the problem, e.g., by freezing the spacetime or continuing the simulation on an axisymmetric grid. The latter requires either a handoff procedure for the NR output (cf. a BHNS merger study with such a procedure~\cite{Gottlieb:2023est}) or the reproduction of the disk using analytical models to construct the initial data.

Here, we investigate the disk structure, compare it with commonly employed analytical prescriptions, and verify their assumptions.

\begin{figure}[tp]
    \centering
    \includegraphics[width=1\linewidth]{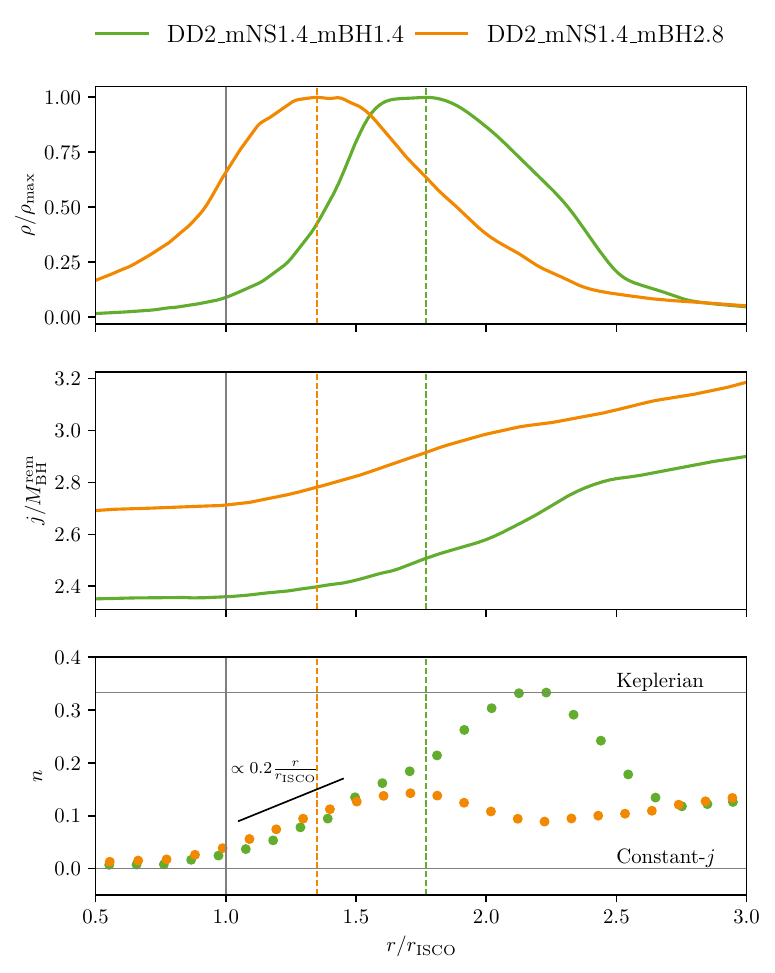}
    \caption{Azimuthal angle- and time-averaged radial profiles of the normalized disk density, specific angular momentum $j$ normalized to the remnant BH mass, and the inferred rotation law power $n$. 
    The gray horizontal lines in the last panel delineate constant-$j$ ($n=0$) and Keplerian ($n=1/3$) regimes.
    The solid gray vertical line shows the ISCO radius, and the dashed vertical lines show the radius at the maximum density of the disk.
    }
    \label{fig:rotation_law}
\end{figure}

\begin{figure}[tp]
    \centering
    \includegraphics[width=1\linewidth]{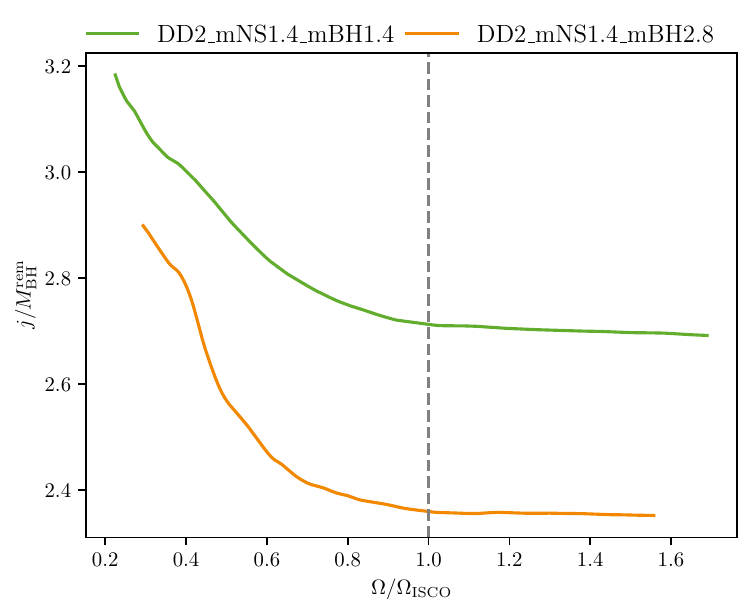}
    \caption{Specific angular momentum $j$ as a function of the angular velocity $\Omega$ normalized to the angular velocity at the ISCO $\Omega_\mathrm{ISCO}$. Note that the values of $\Omega > \Omega_\mathrm{ISCO}$ correspond to the matter within the ISCO.}
    \label{fig:rotation_law_jOmega}
\end{figure}

\subsection{Qualitative dynamics}
We select two representative cases with the sole difference in mass ratio, \verb|DD2_mNS1.4_mBH1.4| and \verb|DD2_mNS1.4_mBH2.8|.
To highlight the differences in morphology of these disks, Fig.~\ref{fig:disk_over_time} shows the evolution of their densities in the equatorial plane over time. Qualitatively, the main deciding factor for the fate and type of the accretion disk is the mass ratio of the binary. In the equal mass ratio case, the disk forms directly from the disrupted material of the NS, and becomes axisymmetric almost immediately, at $\sim 3$~ms after merger. 
On the other hand, in the case of a mass ratio $1/2$, the axisymmetric structure of the disk is disrupted by the interacting spiral density waves in the disk, as well as by the fallback of the massive disruption tail onto the disk.

\subsection{Rotation law}

Throughout the evolution of these disks, we calculate the specific angular momentum $j = h u_\phi$ and angular velocity $\Omega = u^\phi / u^t$, both with the coordinate origin at the location of the BH.
For our remnant systems, we adopt the following power-law rotation law~\cite{Fujibayashi:2020qda} (cf. Ref.~\cite{Karkowski:2018mxf} for a more general rotation law):
\begin{equation}
    \label{eq:rotational_law}
    j = A_n \Omega^{-n},
\end{equation}
with $A_n$ and $n$ being free parameters.
In Fig.~\ref{fig:rotation_law}, we show the radial profiles of the normalized disk density and specific angular momentum $j$ in the equatorial plane averaged over the azimuthal angle and time\footnote{Time average is performed over $[10,12.6]$~ms after the merger for the $q=1/2$ case, and over $[7.15,7.5]$~ms after the merger for the $q=1$ case.}. In the same figure, we plot the rotation law power $n$ obtained by piecewise fitting of Eq.~\eqref{eq:rotational_law} to the numerical data.

For the disk with $q=1$, the matter located within the innermost stable orbit (ISCO) follows the rotation law of constant $j$ ($n=0$), which is the case due to the lack of an effective angular momentum transfer mechanism, as the matter falls freely onto the BH. Outside the ISCO radius and throughout the disk, the power $n$ grows gradually, reaching approximately the Keplerian value, $n=1/3$, at the outer part of the disk. 
Staring at $\sim 2.2 r_\mathrm{ISCO}$, where the disk density drops to around 75\% of its maximum, the rotation law transitions away from Keplerian value and towards a constant one. This is due to the outer part of the disk being only marginally bound and dynamically infalling, thus having inefficient angular momentum transport leading to a sub-Keplerian rotation law. 

In contrast, the disk in the $q=1/2$ case does not become axisymmetric shortly after merger, in contrast to the $q=1$ case, due to interacting spiral waves and fallback of the tidal tail, both of which disrupt the structure of the disk. At $10$~ms, the disk settles into a more stable and axisymmetric configuration. As for the $q=1$ case, the disk matter within the ISCO has constant specific angular momentum. Starting from ISCO and until the radius of maximum density, the power of the rotation law of the disk grows and practically coincides with that of the $q=1$ case. However, $n$ never reaches the Keplerian value and takes on the maximum value of $0.13$. Once the disk density reduces to around 75\% of its maximum at $\sim 1.6 r_\mathrm{ISCO}$, the rotation law power decreases, and later approaches an almost constant value. This behavior replicates that of the $q=1$ case. The $q=1/2$ disk differs from the $q=1$ one primarily by the larger fraction of matter located within the ISCO and the density peak located closer to the BH.

The observed evolution of the rotation law power for both disks is in contrast with common assumptions of a fixed value of $n$ in long-term axisymmetric simulations of the BHNS postmerger disks, e.g., constant specific angular momentum (Fishbone-Moncrief torus \cite{FM1976}) ~\cite{Siegel:2017nub,Fernandez:2018kax,Fahlman:2022jkh}, or a fixed-exponent rotation law between constant and Keplerian~\cite{Rezzolla:2010fd,Fujibayashi:2020qda}.
We do a linear fit for $n(r/r_\mathrm{ISCO})$ for radii starting from the ISCO until the radius of the maximum density, and find that it follows the slope of roughly $n \propto 0.2~r/r_\mathrm{ISCO}$ on average between the two models.

Nevertheless, while we operate within the framework of general-relativistic hydrodynamics, realistic disks would have nonnegligible magnetic fields and neutrino emission, both of which are capable of substantially altering the rotation law~\cite{Etienne:2011ea,Foucart:2015vpa}. Further simulations with additional physics are required to verify whether the rotation law form persists.

The fact that the rotation law can not be approximated by a simple power-law form suggests a more complex functional form. Therefore, for completeness, we provide the rotation law as the dependency $j=j(\Omega)$ in Fig.~\ref{fig:rotation_law_jOmega}. There, the angular velocity is normalized to its value at the ISCO,

\begin{equation}
    \Omega_\mathrm{ISCO} = \frac{1}{M_\mathrm{BH}^\mathrm{rem}}  \frac{1}{ R_\mathrm{ISCO}(\chi_\mathrm{BH}^\mathrm{rem})^{3/2} + \chi_\mathrm{BH}^\mathrm{rem}},
\end{equation}

where $R_\mathrm{ISCO}$ is the dimentionless mass-rescaled radius of the ISCO in the Bardeen-Press-Teukolsky formulation~\cite{Bardeen:1972fi}.

\subsection{Disk oscillations}

For seven configurations where the disk mass $M_\mathrm{disk}$ is larger than $10^{-2}~\mathrm{M_\odot}$ (except \verb|DD2_mNS0.8_mBH1.6|), we identify global density oscillations of the disk. These oscillations are sustained for the simulation time, except for \verb|DD2_mNS1.2_mBH2.4| and \verb|SLy_mNS1.4_mBH1.4|, where only one cycle is present, followed by their subsequent damping.
We find those oscillations to be trapped $g$-modes~\cite{Okazaki:1987,1992ApJ...393..697N,Perez:1996ti,Wagoner:1998hh,Abramowicz:2011xu,Brink:2015roa}, as verified by comparing the dominant frequencies of the density norm $||\rho||$ with eigenfrequencies of $g$-modes assuming them to be approximately equal to the maximum of prograde epicyclic frequency $\kappa$~\cite{Okazaki:1987}:
\begin{equation}
    \label{eq:epicyclic_frequency}
    \kappa^2 = \frac{M_\text{BH}^{\text{rem}} (r^2 - 6 M_\text{BH}^{\text{rem}} r + 8 a (M_\text{BH}^{\text{rem}})^{1/2} r^{1/2} - 3 a^2) }{r^2 \left(r^{3/2} + a (M_\text{BH}^{\text{rem}})^{1/2}\right)^2 },
\end{equation}
where $M_\text{BH}^{\text{rem}}$ and $a$ are the mass and spin parameter of the remnant BH, and $r$ is the Boyer-Lindquist radial coordinate. We maximize $\kappa$ over the radius $r/M_\text{BH}^{\text{rem}}~\in~[1,30]$ and list the predicted analytical epicyclic frequencies in Tab.~\ref{table:remnant_BH_fits} and find that they differ on average from the measured ones in the simulations by 4\% on average.

However, we note that the $g$-modes might also be dampened in the presence of magnetic fields~\cite{Fu:2008iw,Reynolds:2008wp,Ortega-Rodriguez:2015dpa}. Hence, further simulations of BHNS systems, incorporating magnetic fields, would be necessary to assess their relevance.

If these disk oscillations persist and modulate the accretion rate (cf. next subsection), they might be observable as quasiperiodic oscillations (QPOs) in gamma-ray bursts (GRBs). Within this context, we find the $g$-mode eigenfrequencies for the equal mass-ratio systems with component masses between $1.2~M_\odot$ and $1.4~M_\odot$ are either close or compatible with those of high-frequency QPOs found in GRB~181222B~\cite{Yang:2025jfx} and GRB~931101B~\cite{Chirenti:2023dzl}.

\begin{figure}[!htp]
    \centering
    \includegraphics[width=1\linewidth]{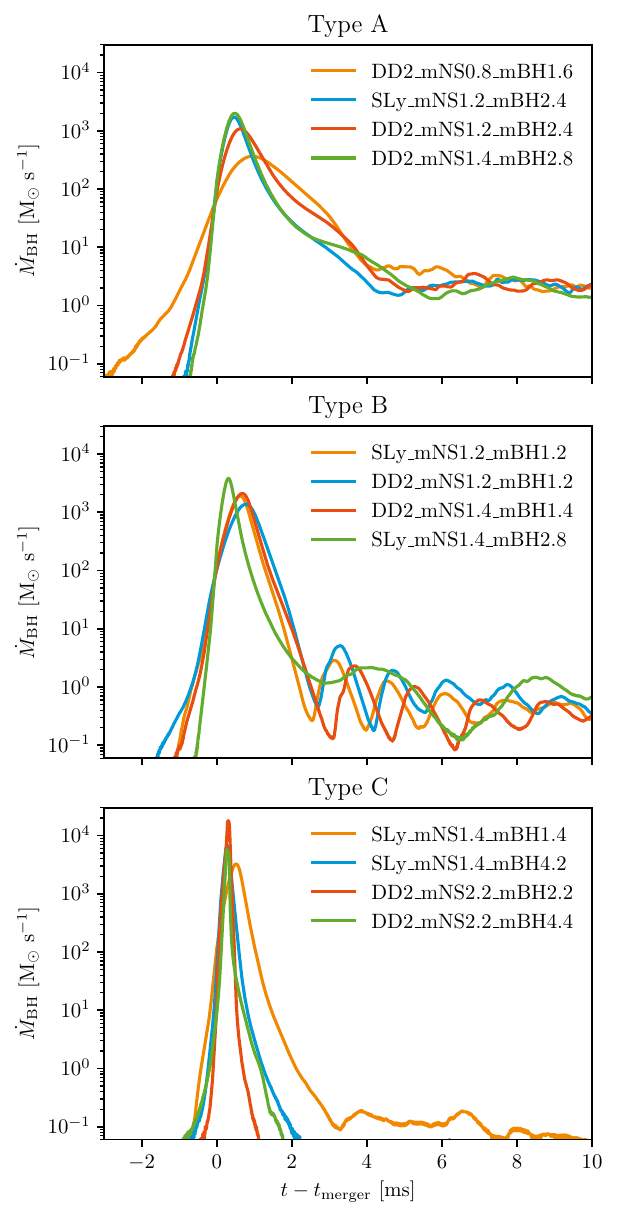}
    \caption{Rate of accretion onto the BH during merger and after merger. The configurations are categorized by merger type: Type A involves disk perturbations by the spiral density waves and the fallback material, Type B involves instant circularization, and Type C involves the NS plunging directly into the BH.}
    \label{fig:accretion}
\end{figure}

\subsection{Accretion}

In Fig.~\ref{fig:accretion}, we show the rate of matter accretion onto the BH over time. We identify three distinct types (A, B, and C) of the accretion rate behavior, and therefore, we plot them separately. For each BHNS configuration, the first accretion peak always includes the mass transfer before the merger and the merger itself, i.e., when the bulk of the NS matter falls onto the BH.  

For the Type~A configurations, a high accretion rate is sustained for $\sim 4$~ms, after which all models approach a steady accretion rate of about $2-3~\mathrm{M_\odot~s^{-1}}$, regardless of their total or NS masses. We note that all Type~A configurations have mass ratio $q=1/2$.

For the Type~B configurations, the initial merger accretion peak ends earlier, at roughly $2$~ms after the merger. After that, the accretion rate shows strong variability, which we find to be directly related to the global oscillations of the disk, where the position of the inner part of the disk periodically contracts, driving accretion by overflowing into the ISCO.
Considering the low accretion rate for Type~B and the order-of-magnitude accretion rate variability, we conclude that the accretion is primarily driven by the $g$-mode oscillations discussed above, which occur for otherwise stable axisymmetric disks.

Finally, for the Type~C configurations, the NS directly plunges into the BH, and the accretion rate peaks only at merger (except \verb|SLy_mNS1.4_mBH1.4|, where only very little matter survives outside the BH). Compared to the merger peaks of Types~A and B, that of Type~C has overall shorter plunging timescales. 

Comparing the disk dynamics and accretion types here with the classification of Ref.~\cite{Kyutoku:2011vz}, we find our Type~A and their Type~III, our Type~B and their Type~I, and our Type~C and their Type~II to be directly related.

\begin{figure*}[tp]
    \centering
    \includegraphics[width=1\linewidth]{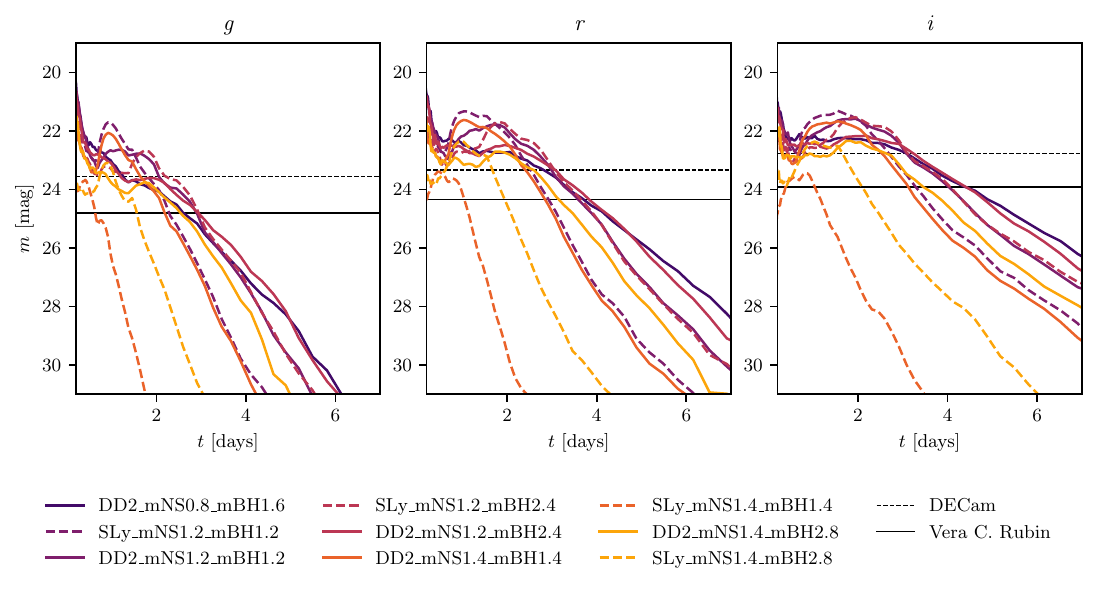}
    \caption{Kilonova lightcurves in $g$, $r$, and $i$ bands for a subset of our BHNS configurations which have enough material for the kilonova emission to not rapidly decay within a day after merger. The lightcurves are given in apparent magnitudes with the source placed at the distance of $200$~Mpc. The black horizontal lines represent the $5\sigma$ depth of the wide-field searches of DECam and Vera C. Rubin Observatory.}
    \label{fig:lightcurves_gri}
\end{figure*}

\section{Kilonova}
\label{section:kilonova}
Using the dynamical ejecta profiles extracted from our simulations at $10$~ms after merger for each system, we perform Monte Carlo radiative transfer simulations using \textsc{POSSIS}~\cite{Bulla:2019muo,Bulla:2022mwo} to estimate properties of the corresponding kilonova emission.  The code uses heating rates from Ref.~\cite{Rosswog:2022tus}, thermalization efficiencies from Ref.~\cite{Wollaeger:2017ahm}, and wavelength- and time-dependent opacities from Ref.~\cite{Tanaka:2019iqp}, which depend on the local properties of the ejecta, including density, temperature, and electron fraction. In \textsc{POSSIS}, the ejecta profiles are rescaled as a function of time, assuming homologous expansion. As the simulations did not include a detailed microphysics of neutrinos, the values for the electron fraction $Y_e$ are inferred using the entropy indicator $\hat{S}$, as described in Refs.~\cite{Neuweiler:2022eum,Markin:2023fxx,Markin:2025oeo}.
For the matter below the $\hat{S}$ threshold, we set the electron fraction to $Y_e=0.15$, and to $Y_e=0.3$ otherwise.
Due to the limited length of our simulations and since we omit the modeling of neutrino emission and magnetic fields, we do not account properly for the wind ejecta, which in our cases would be non-negligible (and often dominant)~\cite{Fernandez:2014bra,Kiuchi:2015qua,Fernandez:2018kax,Nedora:2020hxc} contribution to the kilonova brightness. Therefore, we add a spherical wind ejecta component with a mass of 30\% that of the disk~\cite{Nakar:2019fza}, a constant electron fraction $Y_e = 0.3$, mass-weighted velocity $\bar{v}=0.05$~\cite{Fernandez:2014bra}, and a power-law matter distribution following Ref.~\cite{Bulla:2020jjr}; cf.~Ref.~\cite{Markin:2025oeo} for an application of the same procedure.

We present the resulting lightcurves in SDSS photometric $g$, $r$, and $i$ bands~\cite{Fukugita:1996qt} for the systems that have enough material to sustain kilonova emission over a day after the merger in Fig.~\ref{fig:lightcurves_gri}. The lightcurves are given in apparent magnitudes at the distance of $200$~Mpc, a distance similar to one of GW230529~\cite{LIGOScientific:2024elc}. The lightcurves are primarily dominated by the disk wind ejecta, as evident in the close similarity between \verb|DD2_mNS0.8_mBH1.6| and \verb|DD2_mNS1.2_mBH2.4| cases, which have similar disk masses. All the considered systems have similar peak kilonova brightness around $22$~mag in $r$ and $i$ bands, with differences on the order of $\sim 1$~mag.
The lightcurves are also highly dependent on the EOS: configurations with the softer EOS, SLy, produce a dimmer kilonova than those with a stiffer one, DD2, for the same component masses. The dimmest and rapidly decaying lightcurves are for the configurations with SLy, $M_\mathrm{NS}=1.4$, and both mass ratios, which can be explained by the least total ejecta mass in those systems. Generally, the lightcurves with the brighter peaks later decay faster and become dimmer than those with less bright peaks. This can be explained by the higher optical depth of the more massive wind ejecta.

Additionally, we plot the design $5\sigma$ depths of wide-field surveys of Large Synoptic Survey Telescope (LSST) at Vera C. Rubin Observatory~\cite{LSST:2008ijt} and \mbox{DECam}~\cite{DES:2018gui}. If BHNS mergers were to occur at a distance of $200$~Mpc, it would be possible to locate them (except the two dimmest cases) within the first four days after merger in $r$ and $i$ bands using the Vera~C.~Rubin Observatory. Compared with GW230529, for which no electromagnetic counterparts were found, kilonovae for more symmetric BHNS mergers would be more likely to be located with current wide-field surveys.
 
\section{Conclusion}
\label{section:conclusion}
We performed numerical-relativity simulations of twelve low-mass black hole-neutron star mergers with mass ratios $q\in \{ 1,1/2,1/3\}$, including a system with a subsolar-mass NS.
We do not find good agreement between the BHNS gravitational waveform models and the numerical waveforms from our simulations. In particular, those exhibit dephasing on the order of $1$~rad at the merger for most cases.
For the BHNS system with a subsolar-mass NS, no waveform model operates in this regime of very high tidal deformabilities. Yet, we compare one waveform model with the NS tidal deformability outside of its allowed parameter region, and find large disagreements in both phase and amplitude. Due to a lack of BHNS waveform models for such systems, they will likely be misinterpreted in parameter estimation if they do exist. Overall, updated BHNS waveform models with calibration to a wider range of NR waveforms are needed for accurate interpretation of BHNS systems. 

We analyzed the main merger properties of each system, namely the masses of the dynamical ejecta, disk, and mass and spin of the remnant BH. We compared those values with the corresponding analytical fitting formulas and found good agreement with the mass of the dynamical ejecta and disk, and excellent agreement for the mass of the remnant BH. Yet, while there is a reasonable agreement for the spin of the remnant BH, the updated fitting formula has lower accuracy for spins in systems where the tidal deformability is large. We suggest that this effect might be a result of overfitting for BHNS systems with lower tidal deformabilities and more asymmetric mass ratios, which in turn makes the formula less sensitive to the BH spin-up due to accretion of matter.

We performed a detailed analysis of the dynamics of the accretion disks for two representative cases with $q=\{1,1/2\}$, focusing on their rotation laws and accretion behavior. In line with a previous study of BHNS mergers in the near-equal-mass regime~\cite{Foucart:2019bxj}, these cases represent the formation of differently structured disks: the $q=1$ disk rapidly becomes axisymmetric, while the $q=1/2$ disk starts asymmetric and then settles into an axisymmetric one at around $8$~ms after merger.
For most configurations, we identify global trapped $g$-mode oscillations, the frequency of which is set by the remnant BH.
Most noticeable in the equal-mass-ratio cases, these global density oscillations modulate the accretion rate. For the mass ratio $q=1/2$, the delayed formation of an axisymmetric disk often leads to steady accretion at a universal rate, irrespective of the mass configuration. As these accretion bursts may leave an imprint on the GRB prompt emission, we find the $g$-mode frequencies for some systems compatible with the observed high-frequency QPOs in several GRB lightcurves.

We finalize our analysis by modeling the kilonova emission that these systems would produce. Combining the dynamical ejecta profiles with the spherical wind ejecta model based on our simulations, we perform radiative transfer simulations to obtain the kilonova lightcurves. The lightcurves are primarily dominated by the disk wind ejecta, and thus highly dependent on the EOS employed--the softer EOS produces less massive disks and thus dimmer kilonovae. All lightcurves peak at similar magnitudes around two days after the merger and can last up to six days after the merger for configurations with massive disks. If such BHNS mergers occur at a distance of $200$~Mpc, it would be possible to locate them using the Vera C. Rubin Observatory within the first four days after merger, and by DECam within the first two days after merger. 

Additionally, in the Appendix, we perform a convergence study for the numerical waveforms, disk mass, and remnant BH mass and spin. 
Finally, we correct the numerical waveforms for the drift of the center of mass, which, to the best of our knowledge, has not been applied to BHNS systems in finite-difference NR codes employing moving punctures and general-relativistic hydrodynamics before.

\section{Data availability}
\label{section:data}
The data that support the findings of this article are openly available~\cite{markin_2026_18374717,markin_2026_18377211}. The gravitational waveforms will also be available as part of the CoRe database~\cite{Dietrich:2018phi,Gonzalez:2022mgo}, cf.~Tab.~\ref{table:configurations} for the identifiers.

\section{Acknowledgements}
We are grateful to Adrian G. Abac, Felip A. Ramis Vidal, and Marta Colleoni for their invaluable insights on waveform modeling; to Francois Foucart, Rahime Matur, and Alejandra Gonzalez for useful discussions; and to Maximiliano Ujevic for comments on the waveforms. I.~M. thanks Café Schaubühne for their hospitality.
This manuscript and the supporting software were written without the use of large language models.
I.~M. and T.~D. gratefully acknowledge support from the Deutsche Forschungsgemeinschaft (DFG) under the project 504148597 (DI 2553/7). M.~B. acknowledges the Department of Physics and Earth Science of the University of Ferrara for the financial support through the FIRD 2024 and FIRD 2025 grants. T.~D. acknowledges funding from the European Union (ERC, SMArt, 101076369). 
Views and opinions expressed are those of the authors only and do not necessarily reflect those of the European Union or the European Research Council. Neither the European Union nor the granting authority can be held responsible for them. 
The authors gratefully acknowledge the computing time granted by the Resource Allocation Board and provided on the supercomputer Emmy at NHR-Nord@Göttingen as part of the NHR infrastructure. Most calculations for this research were conducted with computing resources under the project bbp00049. In addition, we acknowledge the usage of the DFG-funded research cluster Jarvis at the University of Potsdam (INST 336/173-1; project number: 502227537).

\appendix

\begin{table*}[!htp]
    \centering
    \begin{tabular}{c|c|ccc|ccc|ccc|c}
Name & $L_\mathrm{BH}^\mathrm{add}$  & $h_0^{R96}$ & $h_{\mathrm{NS}}^\mathrm{R96}$ & $h_\mathrm{finest}^\mathrm{R96}$ & $h_0^{R144}$ & $h_{\mathrm{NS}}^\mathrm{R144}$ & $h_\mathrm{finest}^\mathrm{R144}$ & $h_0^{R192}$ & $h_{\mathrm{NS}}^\mathrm{R192}$ & $h_\mathrm{finest}^\mathrm{R192}$& $R_0$ \\
\hline
\verb|DD2_mNS0.8_mBH1.6| & 1 & 24.832 & $0.194$ & 0.097  & 16.555 & $0.129$ & 0.065  & 12.416 & $0.097$ & 0.048 &  2384 \\
\verb|DD2_mNS1.2_mBH1.2| & 2 & 24.917 & $0.195$ & 0.049  & 16.612 & $0.130$ & 0.032  & 12.459 & $0.097$ & 0.024 &  2392 \\
\verb|DD2_mNS1.2_mBH2.4| & 1 & 24.491 & $0.383^*$ & 0.096  & 16.327 & $0.128$ & 0.064  & 12.245 & $0.096$ & 0.048 &  2351 \\
\verb|DD2_mNS1.4_mBH1.4| & 2 & 24.320 & $0.190$ & 0.047  & 16.213 & $0.127$ & 0.032  & 12.160 & $0.095$ & 0.024 &  2335 \\
\verb|DD2_mNS1.4_mBH2.8| & 1 & 23.893 & $0.187$ & 0.093  & 15.929 & $0.124$ & 0.062  & 11.947 & $0.093$ & 0.047 &  2294 \\
\verb|DD2_mNS2.2_mBH2.2| & 1 & 20.992 & $0.164$ & 0.082  & 13.995 & $0.109$ & 0.055  & 10.496 & $0.082$ & 0.041 &  2015 \\
\verb|DD2_mNS2.2_mBH4.4| & 0 & 20.651 & $0.161$ & 0.161  & 13.767 & $0.108$ & 0.108  & 10.325 & $0.081$ & 0.081 &  1982 \\
\verb|SLy_mNS1.2_mBH1.2| & 2 & 21.333 & $0.167$ & 0.042  & 14.222 & $0.111$ & 0.028  & 10.667 & $0.083$ & 0.021 &  2048 \\
\verb|SLy_mNS1.2_mBH2.4| & 1 & 20.907 & $0.163$ & 0.082  & 13.938 & $0.109$ & 0.054  & 10.453 & $0.082$ & 0.041 &  2007 \\
\verb|SLy_mNS1.4_mBH1.4| & 1 & 20.651 & $0.161$ & 0.081  & 13.767 & $0.108$ & 0.054  & 10.325 & $0.081$ & 0.040 &  1982 \\
\verb|SLy_mNS1.4_mBH2.8| & 0 & 20.224 & $0.158$ & 0.158  & 13.483 & $0.105$ & 0.105  & 10.112 & $0.079$ & 0.079 &  1942 \\
\verb|SLy_mNS1.4_mBH4.2| & 0 & 20.053 & $0.157$ & 0.157  & 13.369 & $0.104$ & 0.104  & 10.027 & $0.078$ & 0.078 &  1925 \\
    \end{tabular}
    \caption{The numerical details of the grid. The first column represents the configuration name, followed by the number of additional refinement levels around the BH $L_\mathrm{BH}^\mathrm{add}$, the grid spacing $h_0$ of the coarsest level, the resolution $h_\mathrm{NS}$ at the finest refinement covering the NS, and at the finest level covering the BH $h_\mathrm{finest}$ for each resolution, and the radius of the outermost level $R_0$. The low-resolution run for DD2\_mNS1.2\_mBH2.4 has the NS covered with only seven levels instead of eight, making the resolution two times coarser. The corresponding grid spacing is marked with an asterisk. }
    \label{tab:numerical_setup}
\end{table*}

\section{Convergence}
\label{section:convergence}

\subsection{Gravitational waveforms}
\begin{figure*}[htp]
    \centering
    \includegraphics[width=1.0\linewidth]{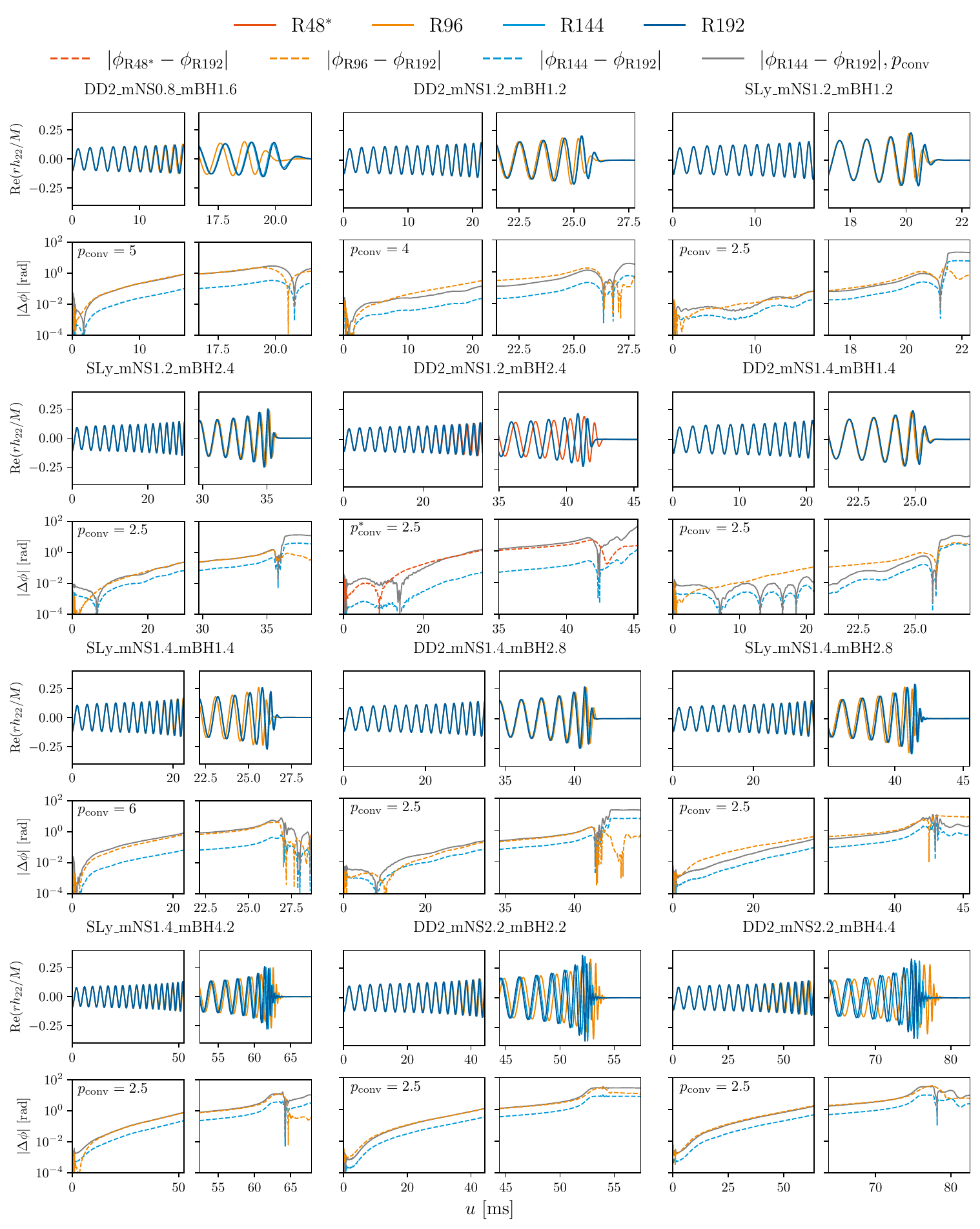}
    \caption{Top panels: The numerical gravitational waveforms at three different resolutions. Bottom panels: Dephasing of the waveforms at low and medium resolution relative to the one at high resolution, and the rescaled dephasing between waveforms at the medium and high resolutions assuming a convergence order $p_\mathrm{conv}$. The low-resolution run of DD2\_mNS1.2\_mBH2.4 has the effective resolution of roughly R$48$, cf.~Tab.~\ref{tab:numerical_setup}. We use this effective resolution to calculate the convergence order and mark these results with an asterisk.}
    \label{figure:waveform_convergence}
\end{figure*}

In Fig.~\ref{figure:waveform_convergence}, we plot the dephasing between the numerical waveforms at low (R96) and medium (R144) resolutions and the high resolution (R192). It is immediately apparent that configurations have diverse convergence behavior: for some configurations, the lower-resolution simulations have a merger earlier than the higher-resolution ones (hereafter left-to-right convergence), while for others, the opposite is true (hereafter right-to-left convergence). For one configuration, \verb|SLy_mNS1.4_mBH2.8|, the waveforms do not form a clear convergence series. The separation between the left-to-right and right-to-left cases appears to be modulated by the total system mass: the right-to-left cases have the total system mass $\gtrsim 3\mathrm{M_\odot}$, including the heaviest equal-mass-ratio system. This observation is in line with the convergence tests in Ref.~\cite{Gonzalez:2025xba}, where the same evolution code and Riemann solver were employed, and the considered configuration exhibits right-to-left convergence (without considering very low resolution). There, however, only a single configuration was checked for convergence, and all the configurations have total masses higher than $3\mathrm{M_\odot}$. The GW230529-like configurations in our previous study~\cite{Markin:2025oeo} also confirm this observation---both BHNS and BNS configurations have total masses higher than $3\mathrm{M_\odot}$ and follow right-to-left convergence.

In addition to the dephasing between different resolutions, in Fig.~\ref{figure:waveform_convergence}, we plot the dephasing between the medium and high resolutions rescaled assuming an apparent convergence order $p_\mathrm{conv}$. For many cases, $p_\mathrm{conv}=2.5$, while for others it can be significantly higher, and reach $p_\mathrm{conv}=6$ for the \verb|SLy_mNS1.4_mBH1.4| case. We stress that such a high apparent convergence order is rather coincidental and is driven by the small phase differences between the waveforms at the high and medium resolutions; further increase in resolution will likely not preserve this apparent convergence order. For the equal-mass configurations, except \verb|DD2_mNS2.2_mBH2.2|, the waveforms for medium and high resolutions are essentially coincident. This fact would likely render the higher-resolution simulations for these cases rather futile in terms of increasing the waveform quality.

Our observations suggest that multiple competing effects are responsible for the convergence behavior, and these effects are highly configuration-dependent. Therefore, it is instrumental to simulate the configurations with a wide range of masses at multiple resolutions.

\begin{table*}[!htp]
    \centering
    \begin{tabular}{c|c|c|c|ccc|ccc}
       Name  & $M_{\mathrm{b,disk}}^{\mathrm{R96}}$ & $M_{\mathrm{b,disk}}^{\mathrm{R144}}$ & $M_{\mathrm{b,disk}}^{\mathrm{R192}}$ & $M_{\mathrm{BH,rem}}^{\mathrm{R96}}$ & $M_{\mathrm{BH,rem}}^{\mathrm{R144}}$ & $M_{\mathrm{BH,rem}}^{\mathrm{R192}}$ &
       $\chi_{\mathrm{BH,rem}}^{\mathrm{R96}}$ & $\chi_{\mathrm{BH,rem}}^{\mathrm{R144}}$ & $\chi_{\mathrm{BH,rem}}^{\mathrm{R192}}$ \\
       \hline
\verb|DD2_mNS0.8_mBH1.6| & $1.20 \times 10^{-1}$ & $1.17 \times 10^{-1}$ & $1.17 \times 10^{-1}$ & 2.204 & 2.214 & 2.215 & 0.609 & 0.610 & 0.611 \\
\verb|DD2_mNS1.2_mBH1.2| & $9.16 \times 10^{-2}$ & $9.92 \times 10^{-2}$ & $9.52 \times 10^{-2}$ & 2.279 & 2.275 & 2.278 & 0.851 & 0.851 & 0.851 \\
\verb|SLy_mNS1.2_mBH1.2| & $4.42 \times 10^{-2}$ & $4.37 \times 10^{-2}$ & $4.24 \times 10^{-2}$ & 2.320 & 2.322 & 2.323 & 0.844 & 0.843 & 0.843 \\
\verb|SLy_mNS1.2_mBH2.4| & $9.71 \times 10^{-2}$ & $9.31 \times 10^{-2}$ & $9.04 \times 10^{-2}$ & $3.440$ & 3.444 & 3.447 & $0.664$ & 0.665 & 0.666 \\
\verb|DD2_mNS1.2_mBH2.4| & $1.49^* \times 10^{-1}$ & $1.27 \times 10^{-1}$ & $1.27 \times 10^{-1}$ & $3.380^*$ & 3.402 & 3.403 & $0.642^*$ & 0.651 & 0.651 \\
\verb|DD2_mNS1.4_mBH1.4| & $3.89 \times 10^{-2}$ & $3.86 \times 10^{-2}$ & $3.92 \times 10^{-2}$ & 2.715 & 2.716 & 2.716 & 0.837 & 0.837 & 0.837 \\
\verb|SLy_mNS1.4_mBH1.4| & $3.70 \times 10^{-3}$ & $2.89 \times 10^{-3}$ & $2.93 \times 10^{-3}$ & 2.726 & 2.734 & 2.735 & 0.817 & 0.813 & 0.813 \\
\verb|DD2_mNS1.4_mBH2.8| & $8.71 \times 10^{-2}$ & $8.43 \times 10^{-2}$ & $8.33 \times 10^{-2}$ & 4.030 & 4.033 & 4.034 & 0.667 & 0.668 & 0.668 \\
\verb|SLy_mNS1.4_mBH2.8| & $1.86 \times 10^{-2}$ & $1.47 \times 10^{-2}$ & $1.32 \times 10^{-2}$ & 4.067 & 4.076 & 4.078 & 0.669 & 0.669 & 0.669 \\
\verb|SLy_mNS1.4_mBH4.2| & $3.57 \times 10^{-3}$ & $1.16 \times 10^{-3}$ & $6.20 \times 10^{-4}$  & 5.455 & 5.453 & 5.453 & 0.552 & 0.551 & 0.551 \\
\verb|DD2_mNS2.2_mBH2.2| & $< 1 \times 10^{-4}$ & $<1 \times 10^{-4}$ & $< 1 \times 10^{-4}$ & 4.227 & 4.228 & 4.228 & 0.717 & 0.717 & 0.717 \\
\verb|DD2_mNS2.2_mBH4.4| & $< 1 \times 10^{-4}$ & $<1 \times 10^{-4}$ & $< 1 \times 10^{-4}$ & 6.353 & 6.352 & 6.352 & 0.627 & 0.627 & 0.627 \\
    \end{tabular}
    \caption{Disk mass $M_{\mathrm{b,disk}}$, remnant BH mass $M_{\mathrm{BH,rem}}$ and spin $\chi_{\mathrm{BH,rem}}$ at three different resolutions extracted at $10$~ms after merger. The effective resolution of the low-resolution run of DD2\_mNS1.2\_mBH2.4 is R48, cf.~\ref{tab:numerical_setup}, and these results are therefore marked with an asterisk. } 
    \label{table:big_convergence_table}
\end{table*}

\subsection{Disk mass}
One of the core quantities of interest for the low-mass BHNS mergers is the disk mass. In Tab.~\ref{table:big_convergence_table}, we list the values of the disk masses for the simulations at three different resolutions extracted at $10$~ms after merger. In most cases, the disk masses form convergent series, and have an error between the medium (R144) and high (R192) resolutions $11\%$ on average. The highest error is for the \verb|SLy_mNS1.4_mBH4.2|, where the disk mass at the medium resolution is $87\%$ off from the value at the high resolution; however, the disk mass is already too low to be properly resolved.

\subsection{BH mass and spin}
In Tab.~\ref{table:big_convergence_table}, we list both the mass and the spin of the remnant BH for each configuration at three resolutions extracted at $10$~ms after merger. For most of the configurations, both quantities form convergent series. For configurations where the NS has high compactness (\verb|DD2_mNS2.2_mBH2.2| and \verb|DD2_mNS2.2_mBH4.4|), the mass and spin practically coincide, which we suggest is caused by the lack of any baryonic remnant, and thus the numerical errors from the evolution of matter.
Generally, the BH mass is higher with increasing resolution, which can be explained by the more accurate localization of the apparent horizon. However, this is not the case of \verb|DD2_mNS1.2_mBH1.2|, where at the R96 resolution, the BH mass is larger than for R192.

\section{Center-of-mass drift correction}
\label{section:com_correction}
In our \textsc{BAM} simulations, similar to many other NR codes\ such as \textsc{SpEC}~\cite{Szilagyi:2015rwa,Woodford:2019tlo,Boyle:2019kee,Scheel:2025jct} and \textsc{LAZEV}~\cite{Healy:2020vre}, the center of mass (COM) of the system exhibits spurious motion. 
This motion can originate from residual linar momentum in the initial data, unresolved junk radiation~\cite{Woodford:2019tlo}, or from the movement of the coordinates themselves~\cite{Sun:2025quw}. As the GWs are extracted at fixed finite coordinate radii, the COM motion leads to a mismatch between the center of the radiative sphere and the coordinate origin. In turn, it induces mode mixing on the GW signal, primarily affecting higher-order, odd-$m$ modes by inducing oscillations to both amplitude and phase~\cite{Woodford:2019tlo}.

\begin{figure*}[!htp]
    \centering
    \includegraphics[width=1\linewidth]{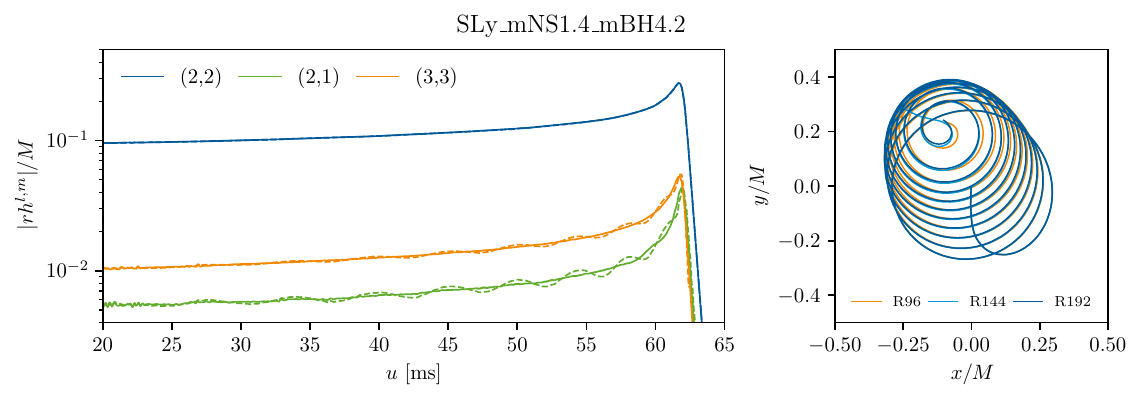}
    \caption{Left: amplitudes of the selected modes before (dashes lines) and after the center-of-mass drift correction (solid lines). Right: the motion of the center of mass in the $x$-$y$ plane starting from the beginning of the simulation, ending at the merger.}
    \label{figure:com}
\end{figure*}

Here, we use the method of Ref.~\cite{Woodford:2019tlo} as implemented in \textsc{scri}~\cite{boyle_2025_17080831} to correct our waveforms for the COM drift using the tracks of the BH puncture and the center of NS. The results are shown in Fig.~\ref{figure:com}, where we highlight the changes in the dominant (2,2) mode, and the two most affected odd-$m$ modes. The COM correction is able to substantially suppress the spurious oscillations in the waveform amplitudes. For completeness, we publish both original and COM-movement-corrected waveforms, cf.~Sect.~\ref{section:data}.

Additionally, in Fig.~\ref{figure:com}, we show the COM tracks at different resolutions. In contrast to the random COM drift directions in \textsc{SpEC}~\cite{Woodford:2019tlo}, our COM tracks appear to form convergent series and follow similar paths across resolutions. This suggests that the COM drift is caused by the junk radiation and that it might become underresolved within the adaptive grid of the spectral code, therefore changing the COM drift magnitude and direction.

\section{Carbon footprint}
\label{section:carbon_footprint}
Performing high-resolution NR simulations requires a large amount of energy. Most of the time, computing facilities are attached to national grids, which acquire energy from a mix of different sources, including fossil fuels. This in turn induces a tangible amount of greenhouse gas (GHG) emissions, which are driving climate change~\cite{Oreskes_2004,Doran_2009,Cook_2013,Cook_2016,Lynas_2021,Myers_2021}.

Here, we quantify the GHG emissions footprint of this study. During runtime, we collect CPU utilization metrics using \textsc{calcium}~\cite{calcium,calcium_zenodo}, which automatically obtains the Thermal Design Power (TDP) of the CPU model, and estimates the energy consumed by the computations.

We list the energy consumption and the corresponding induced greenhouse gas emissions in Table~\ref{table:carbon_footprint}. The energy estimates only account for CPU usage, not that of memory and network components. To account for cooling and other operational overheads, we adopt Power Usage Effectiveness (PUE)~\cite{YUVENTI201390} with a value of $1.03$ for Emmy~\cite{NHRGoettingenPUE} and rescale the CPU energy and emission data accordingly.

However, it is essential to note the transition of the Emmy supercomputer at NHR-Nord@Göttingen to electricity sourced from renewable sources~\cite{NHRGoettingenPUE}, which effectively negates the induced grid emissions.

\begin{table}[!htbp]
    \centering
    \begin{tabular}{l|c|c}
        & Energy & Induced grid emissions,\\
        & MWh & t\ch{CO2}e \\
        \hline
Evolution R192 & 40.6 & 14.0 \\
Evolution R144 & 28.0 & 9.6 \\
Evolution R96 & 6.3 & 2.2 \\
Eccentricity reduction & 6.8 & 2.3 \\
\hline
Total & 81.7 & 28.1 \\
    \end{tabular}
    \caption{Energy consumption for each step of this study and the corresponding estimated induced \ch{CO2}e emissions.}
    \label{table:carbon_footprint}
\end{table}

\bibliography{main}

@article{Typel:2009sy,
    author = "Typel, S. and Ropke, G. and Klahn, T. and Blaschke, D. and Wolter, H. H.",
    title = "{Composition and thermodynamics of nuclear matter with light clusters}",
    eprint = "0908.2344",
    archivePrefix = "arXiv",
    primaryClass = "nucl-th",
    doi = "10.1103/PhysRevC.81.015803",
    journal = "Phys. Rev. C",
    volume = "81",
    pages = "015803",
    year = "2010"
}

@article{Coughlin:2018fis,
    author = "Coughlin, Michael W. and Dietrich, Tim and Margalit, Ben and Metzger, Brian D.",
    title = "{Multimessenger Bayesian parameter inference of a binary neutron star merger}",
    eprint = "1812.04803",
    archivePrefix = "arXiv",
    primaryClass = "astro-ph.HE",
    doi = "10.1093/mnrasl/slz133",
    journal = "Mon. Not. Roy. Astron. Soc.",
    volume = "489",
    number = "1",
    pages = "L91--L96",
    year = "2019"
}

@article{Read:2008iy,
    author = "Read, Jocelyn S. and Lackey, Benjamin D. and Owen, Benjamin J. and Friedman, John L.",
    title = "{Constraints on a phenomenologically parameterized neutron-star equation of state}",
    eprint = "0812.2163",
    archivePrefix = "arXiv",
    primaryClass = "astro-ph",
    doi = "10.1103/PhysRevD.79.124032",
    journal = "Phys. Rev. D",
    volume = "79",
    pages = "124032",
    year = "2009"
}

@article{LIGOScientific:2024elc,
    author = "Abac, A. G. and others",
    collaboration = "LIGO Scientific, Virgo,, KAGRA, VIRGO",
    title = "{Observation of Gravitational Waves from the Coalescence of a 2.5\textendash{}4.5 M$_{\odot}$ Compact Object and a Neutron Star}",
    eprint = "2404.04248",
    archivePrefix = "arXiv",
    primaryClass = "astro-ph.HE",
    reportNumber = "LIGO-P2300352",
    doi = "10.3847/2041-8213/ad5beb",
    journal = "Astrophys. J. Lett.",
    volume = "970",
    number = "2",
    pages = "L34",
    year = "2024"
}

@article{LIGOScientific:2021qlt,
    author = "Abbott, R. and others",
    collaboration = "LIGO Scientific, KAGRA, VIRGO",
    title = "{Observation of Gravitational Waves from Two Neutron Star\textendash{}Black Hole Coalescences}",
    eprint = "2106.15163",
    archivePrefix = "arXiv",
    primaryClass = "astro-ph.HE",
    reportNumber = "LIGO Document P2000357",
    doi = "10.3847/2041-8213/ac082e",
    journal = "Astrophys. J. Lett.",
    volume = "915",
    number = "1",
    pages = "L5",
    year = "2021"
}

@article{Fukugita:1996qt,
    author = "Fukugita, M. and Ichikawa, T. and Gunn, J. E. and Doi, M. and Shimasaku, K. and Schneider, D. P.",
    title = "{The Sloan digital sky survey photometric system}",
    reportNumber = "IASSNS-AST-96-3",
    doi = "10.1086/117915",
    journal = "Astron. J.",
    volume = "111",
    pages = "1748",
    year = "1996"
}

@article{Papenfort:2021hod,
    author = "Papenfort, L. Jens and Tootle, Samuel D. and Grandcl\'ement, Philippe and Most, Elias R. and Rezzolla, Luciano",
    title = "{New public code for initial data of unequal-mass, spinning compact-object binaries}",
    eprint = "2103.09911",
    archivePrefix = "arXiv",
    primaryClass = "gr-qc",
    doi = "10.1103/PhysRevD.104.024057",
    journal = "Phys. Rev. D",
    volume = "104",
    number = "2",
    pages = "024057",
    year = "2021"
}

@article{Grandclement:2009ju,
    author = "Grandclement, Philippe",
    title = "{Kadath: A Spectral solver for theoretical physics}",
    eprint = "0909.1228",
    archivePrefix = "arXiv",
    primaryClass = "gr-qc",
    doi = "10.1016/j.jcp.2010.01.005",
    journal = "J. Comput. Phys.",
    volume = "229",
    pages = "3334--3357",
    year = "2010"
}

@article{York:1998hy,
    author = "York, Jr., James W.",
    title = "{Conformal 'thin sandwich' data for the initial-value problem}",
    eprint = "gr-qc/9810051",
    archivePrefix = "arXiv",
    reportNumber = "IFP-UNC-527",
    doi = "10.1103/PhysRevLett.82.1350",
    journal = "Phys. Rev. Lett.",
    volume = "82",
    pages = "1350--1353",
    year = "1999"
}

@article{Pfeiffer:2002iy,
    author = "Pfeiffer, Harald P. and York, Jr., James W.",
    title = "{Extrinsic curvature and the Einstein constraints}",
    eprint = "gr-qc/0207095",
    archivePrefix = "arXiv",
    doi = "10.1103/PhysRevD.67.044022",
    journal = "Phys. Rev. D",
    volume = "67",
    pages = "044022",
    year = "2003"
}

@article{Bruegmann:2006ulg,
    author = "Bruegmann, Bernd and Gonzalez, Jose A. and Hannam, Mark and Husa, Sascha and Sperhake, Ulrich and Tichy, Wolfgang",
    title = "{Calibration of Moving Puncture Simulations}",
    eprint = "gr-qc/0610128",
    archivePrefix = "arXiv",
    doi = "10.1103/PhysRevD.77.024027",
    journal = "Phys. Rev. D",
    volume = "77",
    pages = "024027",
    year = "2008"
}

@article{Thierfelder:2011yi,
    author = "Thierfelder, Marcus and Bernuzzi, Sebastiano and Bruegmann, Bernd",
    title = "{Numerical relativity simulations of binary neutron stars}",
    eprint = "1104.4751",
    archivePrefix = "arXiv",
    primaryClass = "gr-qc",
    doi = "10.1103/PhysRevD.84.044012",
    journal = "Phys. Rev. D",
    volume = "84",
    pages = "044012",
    year = "2011"
}

@article{Dietrich:2015iva,
    author = {Dietrich, Tim and Bernuzzi, Sebastiano and Ujevic, Maximiliano and Br\"ugmann, Bernd},
    title = "{Numerical relativity simulations of neutron star merger remnants using conservative mesh refinement}",
    eprint = "1504.01266",
    archivePrefix = "arXiv",
    primaryClass = "gr-qc",
    doi = "10.1103/PhysRevD.91.124041",
    journal = "Phys. Rev. D",
    volume = "91",
    number = "12",
    pages = "124041",
    year = "2015"
}

@article{Bernuzzi:2016pie,
    author = "Bernuzzi, Sebastiano and Dietrich, Tim",
    title = "{Gravitational waveforms from binary neutron star mergers with high-order weighted-essentially-nonoscillatory schemes in numerical relativity}",
    eprint = "1604.07999",
    archivePrefix = "arXiv",
    primaryClass = "gr-qc",
    doi = "10.1103/PhysRevD.94.064062",
    journal = "Phys. Rev. D",
    volume = "94",
    number = "6",
    pages = "064062",
    year = "2016"
}

@article{Dietrich:2018phi,
    author = {Dietrich, Tim and Radice, David and Bernuzzi, Sebastiano and Zappa, Francesco and Perego, Albino and Br\"ugmann, Bernd and Chaurasia, Swami Vivekanandji and Dudi, Reetika and Tichy, Wolfgang and Ujevic, Maximiliano},
    title = "{CoRe database of binary neutron star merger waveforms}",
    eprint = "1806.01625",
    archivePrefix = "arXiv",
    primaryClass = "gr-qc",
    doi = "10.1088/1361-6382/aaebc0",
    journal = "Class. Quant. Grav.",
    volume = "35",
    number = "24",
    pages = "24LT01",
    year = "2018"
}

@article{Campanelli:2005dd,
    author = "Campanelli, Manuela and Lousto, C. O. and Marronetti, P. and Zlochower, Y.",
    title = "{Accurate evolutions of orbiting black-hole binaries without excision}",
    eprint = "gr-qc/0511048",
    archivePrefix = "arXiv",
    doi = "10.1103/PhysRevLett.96.111101",
    journal = "Phys. Rev. Lett.",
    volume = "96",
    pages = "111101",
    year = "2006"
}

@article{Baker:2005vv,
    author = "Baker, John G. and Centrella, Joan and Choi, Dae-Il and Koppitz, Michael and van Meter, James",
    title = "{Gravitational wave extraction from an inspiraling configuration of merging black holes}",
    eprint = "gr-qc/0511103",
    archivePrefix = "arXiv",
    doi = "10.1103/PhysRevLett.96.111102",
    journal = "Phys. Rev. Lett.",
    volume = "96",
    pages = "111102",
    year = "2006"
}

@article{Hilditch:2012fp,
    author = "Hilditch, David and Bernuzzi, Sebastiano and Thierfelder, Marcus and Cao, Zhoujian and Tichy, Wolfgang and Bruegmann, Bernd",
    title = "{Compact binary evolutions with the Z4c formulation}",
    eprint = "1212.2901",
    archivePrefix = "arXiv",
    primaryClass = "gr-qc",
    doi = "10.1103/PhysRevD.88.084057",
    journal = "Phys. Rev. D",
    volume = "88",
    pages = "084057",
    year = "2013"
}

@misc{calcium,
  author = {Ivan Markin},
  title = {Calcium: Tracking carbon footprint of computing},
  year = {2026},
  howpublished = {\url{https://codeberg.org/unkaktus/calcium}},
  note = {Version v1.10.0}
}

@software{calcium_zenodo,
  author       = {Ivan Markin},
  title        = {unkaktus/calcium: Release v1.7.1},
  month        = oct,
  year         = 2025,
  publisher    = {Zenodo},
  version      = {v1.7.1},
  doi          = {10.5281/zenodo.17338858},
  url          = {https://doi.org/10.5281/zenodo.17338858},
  swhid        = {swh:1:dir:8a551c73f1ac17149a5f7068255f0f63643dccb3
                   ;origin=https://doi.org/10.5281/zenodo.13876574;vi
                   sit=swh:1:snp:90eb55f0aa2caae95ee09404411637787823
                   8933;anchor=swh:1:rel:bde9becd881210a0c97a4d69d11f
                   1dc84a7baf4e;path=unkaktus-calcium-007fd40
                  },
}

@article{Thompson:2020nei,
    author = "Thompson, Jonathan E. and Fauchon-Jones, Edward and Khan, Sebastian and Nitoglia, Elisa and Pannarale, Francesco and Dietrich, Tim and Hannam, Mark",
    title = "{Modeling the gravitational wave signature of neutron star black hole coalescences}",
    eprint = "2002.08383",
    archivePrefix = "arXiv",
    primaryClass = "gr-qc",
    reportNumber = "LIGO-P2000059",
    doi = "10.1103/PhysRevD.101.124059",
    journal = "Phys. Rev. D",
    volume = "101",
    number = "12",
    pages = "124059",
    year = "2020"
}

@article{Bulla:2019muo,
    author = "Bulla, Mattia",
    title = "{POSSIS: predicting spectra, light curves and polarization for multi-dimensional models of supernovae and kilonovae}",
    eprint = "1906.04205",
    archivePrefix = "arXiv",
    primaryClass = "astro-ph.HE",
    doi = "10.1093/mnras/stz2495",
    journal = "Mon. Not. Roy. Astron. Soc.",
    volume = "489",
    number = "4",
    pages = "5037--5045",
    year = "2019"
}

@article{Bulla:2022mwo,
    author = "Bulla, Mattia",
    title = "{The critical role of nuclear heating rates, thermalization efficiencies, and opacities for kilonova modelling and parameter inference}",
    eprint = "2211.14348",
    archivePrefix = "arXiv",
    primaryClass = "astro-ph.HE",
    doi = "10.1093/mnras/stad232",
    journal = "Mon. Not. Roy. Astron. Soc.",
    volume = "520",
    number = "2",
    pages = "2558--2570",
    year = "2023"
}

@article{Tanaka:2019iqp,
    author = "Tanaka, Masaomi and Kato, Daiji and Gaigalas, Gediminas and Kawaguchi, Kyohei",
    title = "{Systematic Opacity Calculations for Kilonovae}",
    eprint = "1906.08914",
    archivePrefix = "arXiv",
    primaryClass = "astro-ph.HE",
    doi = "10.1093/mnras/staa1576",
    journal = "Mon. Not. Roy. Astron. Soc.",
    volume = "496",
    number = "2",
    pages = "1369--1392",
    year = "2020"
}

@article{Wollaeger:2017ahm,
    author = "Wollaeger, Ryan T. and Korobkin, Oleg and Fontes, Christopher J. and Rosswog, Stephan K. and Even, Wesley P. and Fryer, Christopher L. and Sollerman, Jesper and Hungerford, Aimee L. and van Rossum, Daniel R. and Wollaber, Allan B.",
    title = "{Impact of ejecta morphology and composition on the electromagnetic signatures of neutron star mergers}",
    eprint = "1705.07084",
    archivePrefix = "arXiv",
    primaryClass = "astro-ph.HE",
    reportNumber = "LA-UR-17-24109",
    doi = "10.1093/mnras/sty1018",
    journal = "Mon. Not. Roy. Astron. Soc.",
    volume = "478",
    number = "3",
    pages = "3298--3334",
    year = "2018"
}

@article{Rosswog:2022tus,
    author = "Rosswog, Stephan and Korobkin, Oleg",
    title = "{Heavy Elements and Electromagnetic Transients from Neutron Star Mergers}",
    eprint = "2208.14026",
    archivePrefix = "arXiv",
    primaryClass = "astro-ph.HE",
    doi = "10.1002/andp.202200306",
    journal = "Annalen Phys.",
    volume = "536",
    number = "2",
    pages = "2200306",
    year = "2024"
}

@article{Markin:2023fxx,
    author = "Markin, Ivan and Neuweiler, Anna and Abac, Adrian and Chaurasia, Swami Vivekanandji and Ujevic, Maximiliano and Bulla, Mattia and Dietrich, Tim",
    title = "{General-relativistic hydrodynamics simulation of a neutron star\textendash{}sub-solar-mass black hole merger}",
    eprint = "2304.11642",
    archivePrefix = "arXiv",
    primaryClass = "gr-qc",
    doi = "10.1103/PhysRevD.108.064025",
    journal = "Phys. Rev. D",
    volume = "108",
    number = "6",
    pages = "064025",
    year = "2023"
}

@article{YUVENTI201390,
title = {A critical analysis of Power Usage Effectiveness and its use in communicating data center energy consumption},
journal = {Energy and Buildings},
volume = {64},
pages = {90-94},
year = {2013},
issn = {0378-7788},
doi = {https://doi.org/10.1016/j.enbuild.2013.04.015},
url = {https://www.sciencedirect.com/science/article/pii/S037877881300251X},
author = {Jumie Yuventi and Roshan Mehdizadeh},
}

@article{Myers_2021,
doi = {10.1088/1748-9326/ac2774},
url = {https://dx.doi.org/10.1088/1748-9326/ac2774},
year = {2021},
month = {oct},
publisher = {IOP Publishing},
volume = {16},
number = {10},
pages = {104030},
author = {Myers, Krista F and Doran, Peter T and Cook, John and Kotcher, John E and Myers, Teresa A},
title = {Consensus revisited: quantifying scientific agreement on climate change and climate expertise among Earth scientists 10 years later},
journal = {Environmental Research Letters}
}

@article{Lynas_2021,
	title        = {Greater than 99\% consensus on human caused climate change in the peer-reviewed scientific literature},
	author       = {Lynas, Mark and Houlton, Benjamin Z and Perry, Simon},
	year         = 2021,
	month        = {oct},
	journal      = {Environmental Research Letters},
	publisher    = {IOP Publishing},
	volume       = 16,
	number       = 11,
	pages        = 114005,
	doi          = {10.1088/1748-9326/ac2966},
	url          = {https://dx.doi.org/10.1088/1748-9326/ac2966}
}

@article{Cook_2016,
doi = {10.1088/1748-9326/11/4/048002},
url = {https://dx.doi.org/10.1088/1748-9326/11/4/048002},
year = {2016},
month = {apr},
publisher = {IOP Publishing},
volume = {11},
number = {4},
pages = {048002},
author = {Cook, John and Oreskes, Naomi and Doran, Peter T and Anderegg, William R L and Verheggen, Bart and Maibach, Ed W and Carlton, J Stuart and Lewandowsky, Stephan and Skuce, Andrew G and Green, Sarah A and Nuccitelli, Dana and Jacobs, Peter and Richardson, Mark and Winkler, Bärbel and Painting, Rob and Rice, Ken},
title = {Consensus on consensus: a synthesis of consensus estimates on human-caused global warming},
journal = {Environmental Research Letters}
}

@article{Cook_2013,
doi = {10.1088/1748-9326/8/2/024024},
url = {https://dx.doi.org/10.1088/1748-9326/8/2/024024},
year = {2013},
month = {may},
publisher = {IOP Publishing},
volume = {8},
number = {2},
pages = {024024},
author = {Cook, John and Nuccitelli, Dana and Green, Sarah A and Richardson, Mark and Winkler, Bärbel and Painting, Rob and Way, Robert and Jacobs, Peter and Skuce, Andrew},
title = {Quantifying the consensus on anthropogenic global warming in the scientific literature},
journal = {Environmental Research Letters},
}

@article{Doran_2009,
author = {Doran, Peter T. and Zimmerman, Maggie Kendall},
title = {Examining the Scientific Consensus on Climate Change},
journal = {Eos, Transactions American Geophysical Union},
volume = {90},
number = {3},
pages = {22-23},
doi = {https://doi.org/10.1029/2009EO030002},
year = {2009}
}

@article{Oreskes_2004,
author = {Naomi Oreskes},
title = {The Scientific Consensus on Climate Change},
journal = {Science},
volume = {306},
number = {5702},
pages = {1686-1686},
year = {2004},
doi = {10.1126/science.1103618},
URL = {https://www.science.org/doi/abs/10.1126/science.1103618},
eprint = {https://www.science.org/doi/pdf/10.1126/science.1103618}}

@article{DES:2018gui,
    author = "Abbott, T. M. C. and others",
    collaboration = "DES, NOAO Data Lab",
    title = "{The Dark Energy Survey Data Release 1}",
    eprint = "1801.03181",
    archivePrefix = "arXiv",
    primaryClass = "astro-ph.IM",
    reportNumber = "FERMILAB-PUB-17-603-AE-E",
    doi = "10.3847/1538-4365/aae9f0",
    journal = "Astrophys. J. Suppl.",
    volume = "239",
    number = "2",
    pages = "18",
    year = "2018"
}

@article{LSST:2008ijt,
    author = "Ivezi\'c, \v{Z}eljko and others",
    collaboration = "LSST",
    title = "{LSST: from Science Drivers to Reference Design and Anticipated Data Products}",
    eprint = "0805.2366",
    archivePrefix = "arXiv",
    primaryClass = "astro-ph",
    reportNumber = "SLAC-PUB-16076",
    doi = "10.3847/1538-4357/ab042c",
    journal = "Astrophys. J.",
    volume = "873",
    number = "2",
    pages = "111",
    year = "2019"
}

@article{Kruger:2020gig,
    author = {Kr\"uger, Christian J\"urgen and Foucart, Francois},
    title = "{Estimates for Disk and Ejecta Masses Produced in Compact Binary Mergers}",
    eprint = "2002.07728",
    archivePrefix = "arXiv",
    primaryClass = "astro-ph.HE",
    doi = "10.1103/PhysRevD.101.103002",
    journal = "Phys. Rev. D",
    volume = "101",
    number = "10",
    pages = "103002",
    year = "2020"
}

@ARTICLE{2008JCoPh.227.3191B,
       author = {{Borges}, Rafael and {Carmona}, Monique and {Costa}, Bruno and {Don}, Wai Sun},
        title = "{An improved weighted essentially non-oscillatory scheme for hyperbolic conservation laws}",
      journal = {Journal of Computational Physics},
         year = 2008,
        month = mar,
       volume = {227},
       number = {6},
        pages = {3191-3211},
          doi = {10.1016/j.jcp.2007.11.038},
       adsurl = {https://ui.adsabs.harvard.edu/abs/2008JCoPh.227.3191B}
}

@article{Li:1998bw,
    author = "Li, Li-Xin and Paczynski, Bohdan",
    title = "{Transient events from neutron star mergers}",
    eprint = "astro-ph/9807272",
    archivePrefix = "arXiv",
    reportNumber = "POPE-772",
    doi = "10.1086/311680",
    journal = "Astrophys. J. Lett.",
    volume = "507",
    pages = "L59",
    year = "1998"
}

@article{Kiuchi:2015qua,
    author = "Kiuchi, Kenta and Sekiguchi, Yuichiro and Kyutoku, Koutarou and Shibata, Masaru and Taniguchi, Keisuke and Wada, Tomohide",
    title = "{High resolution magnetohydrodynamic simulation of black hole-neutron star merger: Mass ejection and short gamma ray bursts}",
    eprint = "1506.06811",
    archivePrefix = "arXiv",
    primaryClass = "astro-ph.HE",
    doi = "10.1103/PhysRevD.92.064034",
    journal = "Phys. Rev. D",
    volume = "92",
    number = "6",
    pages = "064034",
    year = "2015"
}

@article{Foucart:2018rjc,
    author = "Foucart, Francois and Hinderer, Tanja and Nissanke, Samaya",
    title = "{Remnant baryon mass in neutron star-black hole mergers: Predictions for binary neutron star mimickers and rapidly spinning black holes}",
    eprint = "1807.00011",
    archivePrefix = "arXiv",
    primaryClass = "astro-ph.HE",
    doi = "10.1103/PhysRevD.98.081501",
    journal = "Phys. Rev. D",
    volume = "98",
    number = "8",
    pages = "081501",
    year = "2018"
}

@article{Varma:2019csw,
    author = "Varma, Vijay and Field, Scott E. and Scheel, Mark A. and Blackman, Jonathan and Gerosa, Davide and Stein, Leo C. and Kidder, Lawrence E. and Pfeiffer, Harald P.",
    title = "{Surrogate models for precessing binary black hole simulations with unequal masses}",
    eprint = "1905.09300",
    archivePrefix = "arXiv",
    primaryClass = "gr-qc",
    doi = "10.1103/PhysRevResearch.1.033015",
    journal = "Phys. Rev. Research.",
    volume = "1",
    pages = "033015",
    year = "2019"
}

@article{Kunnumkai:2024qmw,
    author = "Kunnumkai, Keerthi and Palmese, Antonella and Bulla, Mattia and Dietrich, Tim and Farah, Amanda M. and Pang, Peter T. H.",
    title = "{Kilonova emission from GW230529 and mass gap neutron star-black hole mergers}",
    eprint = "2409.10651",
    archivePrefix = "arXiv",
    primaryClass = "astro-ph.HE",
    doi = "10.1103/dnjl-gc4x",
    journal = "Phys. Rev. D",
    volume = "112",
    number = "12",
    pages = "123005",
    year = "2025"
}

@misc{NHRGoettingenPUE,
  author = {Gesellschaft für wissenschaftliche Datenverarbeitung mbH Göttingen},
  title = {{GWDG setzt auf klimafreundlichen IT-Betrieb}},
  howpublished = "\url{https://gwdg.de/about-us/press-releases/2021/press-release-4-2021/}",
  year = {2020}, 
}

@article{Neuweiler:2022eum,
    author = "Neuweiler, Anna and Dietrich, Tim and Bulla, Mattia and Chaurasia, Swami Vivekanandji and Rosswog, Stephan and Ujevic, Maximiliano",
    title = "{Long-term simulations of dynamical ejecta: Homologous expansion and kilonova properties}",
    eprint = "2208.13460",
    archivePrefix = "arXiv",
    primaryClass = "astro-ph.HE",
    doi = "10.1103/PhysRevD.107.023016",
    journal = "Phys. Rev. D",
    volume = "107",
    number = "2",
    pages = "023016",
    year = "2023"
}

@article{Nedora:2020hxc,
    author = "Nedora, Vsevolod and Bernuzzi, Sebastiano and Radice, David and Daszuta, Boris and Endrizzi, Andrea and Perego, Albino and Prakash, Aviral and Safarzadeh, Mohammadtaher and Schianchi, Federico and Logoteta, Domenico",
    title = "{Numerical Relativity Simulations of the Neutron Star Merger GW170817: Long-Term Remnant Evolutions, Winds, Remnant Disks, and Nucleosynthesis}",
    eprint = "2008.04333",
    archivePrefix = "arXiv",
    primaryClass = "astro-ph.HE",
    doi = "10.3847/1538-4357/abc9be",
    journal = "Astrophys. J.",
    volume = "906",
    number = "2",
    pages = "98",
    year = "2021"
}

@article{Martineau:2024zur,
    author = "Martineau, Tia and Foucart, Francois and Scheel, Mark A. and Duez, Matthew D. and Kidder, Lawrence E. and Pfeiffer, Harald P.",
    title = "{Black hole-neutron star binaries near neutron star disruption limit in the mass regime of event GW230529}",
    eprint = "2405.06819",
    archivePrefix = "arXiv",
    primaryClass = "astro-ph.HE",
    doi = "10.1088/1361-6382/ae2c37",
    journal = "Class. Quant. Grav.",
    volume = "43",
    number = "1",
    pages = "015015",
    year = "2026"
}

@article{Gonzalez:2022prs,
    author = "Gonzalez, Alejandra and Gamba, Rossella and Breschi, Matteo and Zappa, Francesco and Carullo, Gregorio and Bernuzzi, Sebastiano and Nagar, Alessandro",
    title = "{Numerical-relativity-informed effective-one-body model for black-hole{\textendash}neutron-star mergers with higher modes and spin precession}",
    eprint = "2212.03909",
    archivePrefix = "arXiv",
    primaryClass = "gr-qc",
    doi = "10.1103/PhysRevD.107.084026",
    journal = "Phys. Rev. D",
    volume = "107",
    number = "8",
    pages = "084026",
    year = "2023"
}

@article{Fernandez:2014bra,
    author = "Fern{\'a}ndez, Rodrigo and Quataert, Eliot and Schwab, Josiah and Kasen, Daniel and Rosswog, Stephan",
    title = "{The interplay of disc wind and dynamical ejecta in the aftermath of neutron star{\textendash}black hole mergers}",
    eprint = "1412.5588",
    archivePrefix = "arXiv",
    primaryClass = "astro-ph.HE",
    doi = "10.1093/mnras/stv238",
    journal = "Mon. Not. Roy. Astron. Soc.",
    volume = "449",
    number = "1",
    pages = "390--402",
    year = "2015"
}

@article{Raaijmakers:2021slr,
    author = "Raaijmakers, Geert and others",
    title = "{The Challenges Ahead for Multimessenger Analyses of Gravitational Waves and Kilonova: A Case Study on GW190425}",
    eprint = "2102.11569",
    archivePrefix = "arXiv",
    primaryClass = "astro-ph.HE",
    doi = "10.3847/1538-4357/ac222d",
    journal = "Astrophys. J.",
    volume = "922",
    number = "2",
    pages = "269",
    year = "2021"
}

@article{Gonzalez:2025xba,
    author = "Gonzalez, Alejandra and Bernuzzi, Sebastiano and Rashti, Alireza and Brandoli, Francesco and Gamba, Rossella",
    title = "{Black-hole - neutron-star mergers: new numerical-relativity simulations and multipolar effective-one-body model with spin precession and eccentricity}",
    eprint = "2507.00113",
    archivePrefix = "arXiv",
    primaryClass = "gr-qc",
    month = "6",
    year = "2025",
    journal = ""
}

@article{Ozel:2010su,
    author = "Ozel, Feryal and Psaltis, Dimitrios and Narayan, Ramesh and McClintock, Jeffrey E.",
    title = "{The Black Hole Mass Distribution in the Galaxy}",
    eprint = "1006.2834",
    archivePrefix = "arXiv",
    primaryClass = "astro-ph.GA",
    doi = "10.1088/0004-637X/725/2/1918",
    journal = "Astrophys. J.",
    volume = "725",
    pages = "1918--1927",
    year = "2010"
}

@article{Farr:2010tu,
    author = "Farr, Will M. and Sravan, Niharika and Cantrell, Andrew and Kreidberg, Laura and Bailyn, Charles D. and Mandel, Ilya and Kalogera, Vicky",
    title = "{The Mass Distribution of Stellar-Mass Black Holes}",
    eprint = "1011.1459",
    archivePrefix = "arXiv",
    primaryClass = "astro-ph.GA",
    doi = "10.1088/0004-637X/741/2/103",
    journal = "Astrophys. J.",
    volume = "741",
    pages = "103",
    year = "2011"
}

@article{Nakar:2019fza,
    author = "Nakar, Ehud",
    title = "{The electromagnetic counterparts of compact binary mergers}",
    eprint = "1912.05659",
    archivePrefix = "arXiv",
    primaryClass = "astro-ph.HE",
    doi = "10.1016/j.physrep.2020.08.008",
    journal = "Phys. Rept.",
    volume = "886",
    pages = "1--84",
    year = "2020"
}

@article{Mahapatra:2025agb,
    author = "Mahapatra, Parthapratim and Chattopadhyay, Debatri and Gupta, Anuradha and Antonini, Fabio and Favata, Marc and Sathyaprakash, B. S. and Arun, K. G.",
    title = "{Possible binary neutron star merger history of the primary of GW230529}",
    eprint = "2503.17872",
    archivePrefix = "arXiv",
    primaryClass = "astro-ph.HE",
    reportNumber = "LIGO preprint number P2500111",
    doi = "10.1103/c9l3-gw6w",
    journal = "Phys. Rev. D",
    volume = "111",
    number = "12",
    pages = "123030",
    year = "2025"
}

@article{Bulla:2020jjr,
    author = "Bulla, M. and Kyutoku, K. and Tanaka, M. and Covino, S. and Bruten, J. R. and Matsumoto, T. and Maund, J. R. and Testa, V. and Wiersema, K.",
    title = "{Polarized kilonovae from black hole{\textendash}neutron star mergers}",
    eprint = "2009.07279",
    archivePrefix = "arXiv",
    primaryClass = "astro-ph.HE",
    doi = "10.1093/mnras/staa3796",
    journal = "Mon. Not. Roy. Astron. Soc.",
    volume = "501",
    number = "2",
    pages = "1891--1899",
    year = "2021"
}

@article{Fernandez:2018kax,
    author = "Fern{\'a}ndez, Rodrigo and Tchekhovskoy, Alexander and Quataert, Eliot and Foucart, Francois and Kasen, Daniel",
    title = "{Long-term GRMHD simulations of neutron star merger accretion discs: implications for electromagnetic counterparts}",
    eprint = "1808.00461",
    archivePrefix = "arXiv",
    primaryClass = "astro-ph.HE",
    doi = "10.1093/mnras/sty2932",
    journal = "Mon. Not. Roy. Astron. Soc.",
    volume = "482",
    number = "3",
    pages = "3373--3393",
    year = "2019"
}

@article{Douchin:2001sv,
    author = "Douchin, F. and Haensel, P.",
    title = "{A unified equation of state of dense matter and neutron star structure}",
    eprint = "astro-ph/0111092",
    archivePrefix = "arXiv",
    doi = "10.1051/0004-6361:20011402",
    journal = "Astron. Astrophys.",
    volume = "380",
    pages = "151",
    year = "2001"
}

@article{Matur:2025avh,
    author = "Matur, Rahime and Hawke, Ian and Andersson, Nils",
    title = "{Impact of black hole spin on low-mass black hole-neutron star mergers}",
    eprint = "2508.06341",
    archivePrefix = "arXiv",
    primaryClass = "astro-ph.HE",
    month = "8",
    year = "2025",
    journal = ""
}

@article{Matur:2024nwi,
    author = "Matur, Rahime and Hawke, Ian and Andersson, Nils",
    title = "{Signatures of low-mass black hole{\textendash}neutron star mergers}",
    eprint = "2407.18045",
    archivePrefix = "arXiv",
    primaryClass = "astro-ph.HE",
    doi = "10.1093/mnras/stae2238",
    journal = "Mon. Not. Roy. Astron. Soc.",
    volume = "534",
    number = "3",
    pages = "2894--2903",
    year = "2024"
}

@article{Foucart:2019bxj,
    author = "Foucart, F. and Duez, M. D. and Kidder, L. E. and Nissanke, S. and Pfeiffer, H. P. and Scheel, M. A.",
    title = "{Numerical simulations of neutron star-black hole binaries in the near-equal-mass regime}",
    eprint = "1903.09166",
    archivePrefix = "arXiv",
    primaryClass = "astro-ph.HE",
    doi = "10.1103/PhysRevD.99.103025",
    journal = "Phys. Rev. D",
    volume = "99",
    number = "10",
    pages = "103025",
    year = "2019"
}

@article{Baumgarte:2025syh,
    author = "Baumgarte, Thomas W. and Shapiro, Stuart L.",
    title = "{Can Premature Collapse Form Black Holes in the Upper and Lower Mass Gaps?}",
    eprint = "2509.04574",
    archivePrefix = "arXiv",
    primaryClass = "astro-ph.HE",
    doi = "10.1103/26yd-1mhd",
    journal = "Phys. Rev. Lett.",
    volume = "135",
    number = "19",
    pages = "191401",
    year = "2025"
}

@article{Capela:2013yf,
    author = "Capela, Fabio and Pshirkov, Maxim and Tinyakov, Peter",
    title = "{Constraints on primordial black holes as dark matter candidates from capture by neutron stars}",
    eprint = "1301.4984",
    archivePrefix = "arXiv",
    primaryClass = "astro-ph.CO",
    reportNumber = "ULB-TH-13-01",
    doi = "10.1103/PhysRevD.87.123524",
    journal = "Phys. Rev. D",
    volume = "87",
    number = "12",
    pages = "123524",
    year = "2013"
}

@article{Genolini:2020ejw,
    author = "G{\'e}nolini, Yoann and Serpico, Pasquale and Tinyakov, Peter",
    title = "{Revisiting primordial black hole capture into neutron stars}",
    eprint = "2006.16975",
    archivePrefix = "arXiv",
    primaryClass = "astro-ph.HE",
    doi = "10.1103/PhysRevD.102.083004",
    journal = "Phys. Rev. D",
    volume = "102",
    number = "8",
    pages = "083004",
    year = "2020"
}

@article{Abramowicz:2017zbp,
    author = "Abramowicz, Marek A. and Bejger, Micha{\l} and Wielgus, Maciek",
    title = "{Collisions of neutron stars with primordial black holes as fast radio bursts engines}",
    eprint = "1704.05931",
    archivePrefix = "arXiv",
    primaryClass = "astro-ph.HE",
    doi = "10.3847/1538-4357/aae64a",
    journal = "Astrophys. J.",
    volume = "868",
    number = "1",
    pages = "17",
    year = "2018"
}

@article{East:2019dxt,
    author = "East, William E. and Lehner, Luis",
    title = "{Fate of a neutron star with an endoparasitic black hole and implications for dark matter}",
    eprint = "1909.07968",
    archivePrefix = "arXiv",
    primaryClass = "gr-qc",
    doi = "10.1103/PhysRevD.100.124026",
    journal = "Phys. Rev. D",
    volume = "100",
    number = "12",
    pages = "124026",
    year = "2019"
}

@software{boyle_2025_17080831,
  author       = {Boyle, Michael and
                  Iozzo, Dante and
                  Stein, Leo and
                  Khairnar, Aniket and
                  Rüter, Hannes and
                  Scheel, Mark and
                  Varma, Vijay and
                  Mitman, Keefe},
  title        = {scri},
  month        = sep,
  year         = 2025,
  publisher    = {Zenodo},
  version      = {v2024.0.3},
  doi          = {10.5281/zenodo.17080831},
  url          = {https://doi.org/10.5281/zenodo.17080831},
}

@article{Marti:1991wi,
    author = "Marti, Jose M. and Ibanez, Jose M. and Miralles, Juan A.",
    title = "{Numerical relativistic hydrodynamics: Local characteristic approach}",
    reportNumber = "FTUV-91-10",
    doi = "10.1103/PhysRevD.43.3794",
    journal = "Phys. Rev. D",
    volume = "43",
    pages = "3794--3801",
    year = "1991"
}

@article{Banyuls:1997zz,
    author = "Banyuls, Francesc and Font, Jose A. and Ibanez, Jose M. A. and Marti, Jose M. A. and Miralles, Juan A.",
    title = "{Numerical {3+1} General Relativistic Hydrodynamics: A Local Characteristic Approach}",
    journal = "Astrophys. J.",
    volume = "476",
    pages = "221",
    year = "1997"
}

@article{Anton:2005gi,
    author = "Anton, Luis and Zanotti, Olindo and Miralles, Juan A. and Marti, Jose M. and Ibanez, Jose M. and Font, Jose A. and Pons, Jose A.",
    title = "{Numerical 3+1 general relativistic magnetohydrodynamics: A Local characteristic approach}",
    eprint = "astro-ph/0506063",
    archivePrefix = "arXiv",
    doi = "10.1086/498238",
    journal = "Astrophys. J.",
    volume = "637",
    pages = "296--312",
    year = "2006"
}

@article{Font:2008fka,
    author = "Font, Jose A.",
    title = "{Numerical Hydrodynamics and Magnetohydrodynamics in General Relativity}",
    doi = "10.12942/lrr-2008-7",
    journal = "Living Rev. Rel.",
    volume = "11",
    pages = "7",
    year = "2008"
}

@article{Berger:1989a,
  author	= {{Berger}, M.~J. and {Colella}, P.},
  title	= "{Local adaptive mesh refinement for shock hydrodynamics}",
  journal	= {Journal of Computational Physics},
  year	= 1989,
  month	= may,
  volume	= 82,
  pages	= {64-84},
  doi		= {10.1016/0021-9991(89)90035-1}
}

@article{Markin:2025oeo,
    author = "Markin, Ivan and Puecher, Anna and Bulla, Mattia and Dietrich, Tim",
    title = "{Challenging a binary neutron star merger interpretation of GW230529}",
    eprint = "2508.08750",
    archivePrefix = "arXiv",
    primaryClass = "gr-qc",
    reportNumber = "LIGO-P2500457-v3",
    doi = "10.1103/6mtx-nftm",
    journal = "Phys. Rev. D",
    volume = "113",
    number = "2",
    pages = "024031",
    year = "2026"
}

@article{Richards:2021upu,
    author = "Richards, Chloe B. and Baumgarte, Thomas W. and Shapiro, Stuart L.",
    title = "{Accretion onto a small black hole at the center of a neutron star}",
    eprint = "2102.09574",
    archivePrefix = "arXiv",
    primaryClass = "astro-ph.HE",
    doi = "10.1103/PhysRevD.103.104009",
    journal = "Phys. Rev. D",
    volume = "103",
    number = "10",
    pages = "104009",
    year = "2021"
}

@article{Baumgarte:2024iby,
    author = "Baumgarte, Thomas W. and Shapiro, Stuart L.",
    title = "{Primordial black holes captured by neutron stars: Relativistic point-mass treatment}",
    eprint = "2404.08735",
    archivePrefix = "arXiv",
    primaryClass = "gr-qc",
    doi = "10.1103/PhysRevD.109.123012",
    journal = "Phys. Rev. D",
    volume = "109",
    number = "12",
    pages = "123012",
    year = "2024"
}

@article{Baumgarte:2024ouj,
    author = "Baumgarte, Thomas W. and Shapiro, Stuart L.",
    title = "{Primordial black holes captured by neutron stars: Simulations in general relativity}",
    eprint = "2405.10365",
    archivePrefix = "arXiv",
    primaryClass = "gr-qc",
    doi = "10.1103/PhysRevD.110.023021",
    journal = "Phys. Rev. D",
    volume = "110",
    number = "2",
    pages = "023021",
    year = "2024"
}

@article{Takhistov:2017bpt,
    author = "Takhistov, Volodymyr",
    title = "{Transmuted Gravity Wave Signals from Primordial Black Holes}",
    eprint = "1707.05849",
    archivePrefix = "arXiv",
    primaryClass = "astro-ph.CO",
    doi = "10.1016/j.physletb.2018.05.026",
    journal = "Phys. Lett. B",
    volume = "782",
    pages = "77--82",
    year = "2018"
}

@article{Sasaki:2021iuc,
    author = "Sasaki, Misao and Takhistov, Volodymyr and Vardanyan, Valeri and Zhang, Ying-li",
    title = "{Establishing the Nonprimordial Origin of Black Hole{\textendash}Neutron Star Mergers}",
    eprint = "2110.09509",
    archivePrefix = "arXiv",
    primaryClass = "astro-ph.CO",
    reportNumber = "YITP-21-109, IPMU21-0064",
    doi = "10.3847/1538-4357/ac66da",
    journal = "Astrophys. J.",
    volume = "931",
    number = "1",
    pages = "2",
    year = "2022"
}

@article{Kyutoku:2011vz,
    author = "Kyutoku, Koutarou and Okawa, Hirotada and Shibata, Masaru and Taniguchi, Keisuke",
    title = "{Gravitational waves from spinning black hole-neutron star binaries: dependence on black hole spins and on neutron star equations of state}",
    eprint = "1108.1189",
    archivePrefix = "arXiv",
    primaryClass = "astro-ph.HE",
    doi = "10.1103/PhysRevD.84.064018",
    journal = "Phys. Rev. D",
    volume = "84",
    pages = "064018",
    year = "2011"
}

@article{Matas:2020wab,
    author = "Matas, Andrew and others",
    title = "{Aligned-spin neutron-star{\textendash}black-hole waveform model based on the effective-one-body approach and numerical-relativity simulations}",
    eprint = "2004.10001",
    archivePrefix = "arXiv",
    primaryClass = "gr-qc",
    doi = "10.1103/PhysRevD.102.043023",
    journal = "Phys. Rev. D",
    volume = "102",
    number = "4",
    pages = "043023",
    year = "2020"
}

@article{Kawaguchi:2015bwa,
    author = "Kawaguchi, Kyohei and Kyutoku, Koutarou and Nakano, Hiroyuki and Okawa, Hirotada and Shibata, Masaru and Taniguchi, Keisuke",
    title = "{Black hole-neutron star binary merger: Dependence on black hole spin orientation and equation of state}",
    eprint = "1506.05473",
    archivePrefix = "arXiv",
    primaryClass = "astro-ph.HE",
    doi = "10.1103/PhysRevD.92.024014",
    journal = "Phys. Rev. D",
    volume = "92",
    number = "2",
    pages = "024014",
    year = "2015"
}

@article{Fahlman:2022jkh,
    author = "Fahlman, Steven and Fern{\'a}ndez, Rodrigo",
    title = "{Long-term 3D MHD simulations of black hole accretion discs formed in neutron star mergers}",
    eprint = "2204.03005",
    archivePrefix = "arXiv",
    primaryClass = "astro-ph.HE",
    doi = "10.1093/mnras/stac948",
    journal = "Mon. Not. Roy. Astron. Soc.",
    volume = "513",
    number = "2",
    pages = "2689--2707",
    year = "2022"
}

@article{Hayashi:2020zmn,
    author = "Hayashi, Kota and Kawaguchi, Kyohei and Kiuchi, Kenta and Kyutoku, Koutarou and Shibata, Masaru",
    title = "{Properties of the remnant disk and the dynamical ejecta produced in low-mass black hole-neutron star mergers}",
    eprint = "2010.02563",
    archivePrefix = "arXiv",
    primaryClass = "astro-ph.HE",
    doi = "10.1103/PhysRevD.103.043007",
    journal = "Phys. Rev. D",
    volume = "103",
    number = "4",
    pages = "043007",
    year = "2021"
}

@article{Woodford:2019tlo,
    author = "Woodford, Charles J. and Boyle, Michael and Pfeiffer, Harald P.",
    title = "{Compact Binary Waveform Center-of-Mass Corrections}",
    eprint = "1904.04842",
    archivePrefix = "arXiv",
    primaryClass = "gr-qc",
    doi = "10.1103/PhysRevD.100.124010",
    journal = "Phys. Rev. D",
    volume = "100",
    number = "12",
    pages = "124010",
    year = "2019"
}

@article{Rosswog:2005su,
    author = "Rosswog, Stephan",
    title = "{Mergers of neutron star black hole binaries with small mass ratios: Nucleosynthesis, gamma-ray bursts and electromagnetic transients}",
    eprint = "astro-ph/0508138",
    archivePrefix = "arXiv",
    doi = "10.1086/497062",
    journal = "Astrophys. J.",
    volume = "634",
    pages = "1202",
    year = "2005"
}

@article{Lee:2007js,
    author = "Lee, William H. and Ramirez-Ruiz, Enrico",
    title = "{The Progenitors of Short Gamma-Ray Bursts}",
    eprint = "astro-ph/0701874",
    archivePrefix = "arXiv",
    doi = "10.1088/1367-2630/9/1/017",
    journal = "New J. Phys.",
    volume = "9",
    pages = "17",
    year = "2007"
}

@article{Metzger:2019zeh,
    author = "Metzger, Brian D.",
    title = "{Kilonovae}",
    eprint = "1910.01617",
    archivePrefix = "arXiv",
    primaryClass = "astro-ph.HE",
    doi = "10.1007/s41114-019-0024-0",
    journal = "Living Rev. Rel.",
    volume = "23",
    number = "1",
    pages = "1",
    year = "2020"
}

@article{Narayan:1992iy,
    author = "Narayan, Ramesh and Paczynski, Bohdan and Piran, Tsvi",
    title = "{Gamma-ray bursts as the death throes of massive binary stars}",
    eprint = "astro-ph/9204001",
    archivePrefix = "arXiv",
    reportNumber = "CFA-3396",
    doi = "10.1086/186493",
    journal = "Astrophys. J. Lett.",
    volume = "395",
    pages = "L83--L86",
    year = "1992"
}

@article{Janka:1999qu,
    author = "Janka, H. Th. and Eberl, Th. and Ruffert, M. and Fryer, C. L.",
    title = "{Black hole: Neutron star mergers as central engines of gamma-ray bursts}",
    eprint = "astro-ph/9908290",
    archivePrefix = "arXiv",
    reportNumber = "MPA-1209",
    doi = "10.1086/312397",
    journal = "Astrophys. J. Lett.",
    volume = "527",
    pages = "L39",
    year = "1999"
}

@article{Sun:2025quw,
    author = "Sun, Dongze and Ma, Sizheng and Scheel, Mark A. and Teukolsky, Saul A.",
    title = "{Gauge Boundary conditions to mitigate CoM drift in BBH simulations}",
    eprint = "2510.25465",
    archivePrefix = "arXiv",
    primaryClass = "gr-qc",
    month = "10",
    year = "2025",
    journal = "",
}

@article{Healy:2020vre,
    author = "Healy, James and Lousto, Carlos O.",
    title = "{Third RIT binary black hole simulations catalog}",
    eprint = "2007.07910",
    archivePrefix = "arXiv",
    primaryClass = "gr-qc",
    doi = "10.1103/PhysRevD.102.104018",
    journal = "Phys. Rev. D",
    volume = "102",
    number = "10",
    pages = "104018",
    year = "2020"
}

@article{Szilagyi:2015rwa,
    author = "Szil{\'a}gyi, B{\'e}la and Blackman, Jonathan and Buonanno, Alessandra and Taracchini, Andrea and Pfeiffer, Harald P. and Scheel, Mark A. and Chu, Tony and Kidder, Lawrence E. and Pan, Yi",
    title = "{Approaching the Post-Newtonian Regime with Numerical Relativity: A Compact-Object Binary Simulation Spanning 350 Gravitational-Wave Cycles}",
    eprint = "1502.04953",
    archivePrefix = "arXiv",
    primaryClass = "gr-qc",
    doi = "10.1103/PhysRevLett.115.031102",
    journal = "Phys. Rev. Lett.",
    volume = "115",
    number = "3",
    pages = "031102",
    year = "2015"
}

@article{Scheel:2025jct,
    author = "Scheel, Mark A. and others",
    title = "{The SXS collaboration{\textquoteright}s third catalog of binary black hole simulations}",
    eprint = "2505.13378",
    archivePrefix = "arXiv",
    primaryClass = "gr-qc",
    doi = "10.1088/1361-6382/adfd34",
    journal = "Class. Quant. Grav.",
    volume = "42",
    number = "19",
    pages = "195017",
    year = "2025"
}

@article{Boyle:2019kee,
    author = "Boyle, Michael and others",
    title = "{The SXS Collaboration catalog of binary black hole simulations}",
    eprint = "1904.04831",
    archivePrefix = "arXiv",
    primaryClass = "gr-qc",
    doi = "10.1088/1361-6382/ab34e2",
    journal = "Class. Quant. Grav.",
    volume = "36",
    number = "19",
    pages = "195006",
    year = "2019"
}

@ARTICLE{FM1976,
       author = {{Fishbone}, L.~G. and {Moncrief}, V.},
        title = "{Relativistic fluid disks in orbit around Kerr black holes.}",
      journal = {\apj},
         year = 1976,
        month = aug,
       volume = {207},
        pages = {962-976},
          doi = {10.1086/154565},
       adsurl = {https://ui.adsabs.harvard.edu/abs/1976ApJ...207..962F}
}

@article{Siegel:2017nub,
    author = "Siegel, Daniel M. and Metzger, Brian D.",
    title = "{Three-Dimensional General-Relativistic Magnetohydrodynamic Simulations of Remnant Accretion Disks from Neutron Star Mergers: Outflows and $r$-Process Nucleosynthesis}",
    eprint = "1705.05473",
    archivePrefix = "arXiv",
    primaryClass = "astro-ph.HE",
    doi = "10.1103/PhysRevLett.119.231102",
    journal = "Phys. Rev. Lett.",
    volume = "119",
    number = "23",
    pages = "231102",
    year = "2017"
}

@article{Gottlieb:2023est,
    author = "Gottlieb, Ore and others",
    title = "{Large-scale Evolution of Seconds-long Relativistic Jets from Black Hole{\textendash}Neutron Star Mergers}",
    eprint = "2306.14947",
    archivePrefix = "arXiv",
    primaryClass = "astro-ph.HE",
    doi = "10.3847/2041-8213/aceeff",
    journal = "Astrophys. J. Lett.",
    volume = "954",
    number = "1",
    pages = "L21",
    year = "2023"
}

@article{Karkowski:2018mxf,
    author = "Karkowski, Janusz and Kulczycki, Wojciech and Mach, Patryk and Malec, Edward and Odrzywo{\l}ek, Andrzej and Pir{\'o}g, Micha{\l}",
    title = "{Self-gravitating axially symmetric disks in general-relativistic rotation}",
    eprint = "1802.02848",
    archivePrefix = "arXiv",
    primaryClass = "gr-qc",
    doi = "10.1103/PhysRevD.97.104017",
    journal = "Phys. Rev. D",
    volume = "97",
    number = "10",
    pages = "104017",
    year = "2018"
}

@article{Fujibayashi:2020qda,
    author = "Fujibayashi, Sho and Shibata, Masaru and Wanajo, Shinya and Kiuchi, Kenta and Kyutoku, Koutarou and Sekiguchi, Yuichiro",
    title = "{Mass ejection from disks surrounding a low-mass black hole: Viscous neutrino-radiation hydrodynamics simulation in full general relativity}",
    eprint = "2001.04467",
    archivePrefix = "arXiv",
    primaryClass = "astro-ph.HE",
    doi = "10.1103/PhysRevD.101.083029",
    journal = "Phys. Rev. D",
    volume = "101",
    number = "8",
    pages = "083029",
    year = "2020"
}

@article{Rezzolla:2010fd,
    author = "Rezzolla, Luciano and Baiotti, Luca and Giacomazzo, Bruno and Link, David and Font, Jose A.",
    editor = "Ott, Christian D. and Pethick, C. J. and Rezzolla, Luciano",
    title = "{Accurate evolutions of unequal-mass neutron-star binaries: properties of the torus and short GRB engines}",
    eprint = "1001.3074",
    archivePrefix = "arXiv",
    primaryClass = "gr-qc",
    doi = "10.1088/0264-9381/27/11/114105",
    journal = "Class. Quant. Grav.",
    volume = "27",
    pages = "114105",
    year = "2010"
}

@article{Okazaki:1987,
    author = {Okazaki, Atsuo T. and Kato, Shoji and Fukue, Jun},
    title = {Global Trapped Oscillations of Relativistic Accretion Disks},
    journal = {Publications of the Astronomical Society of Japan},
    volume = {39},
    number = {3},
    pages = {457-473},
    year = {1987},
    month = {08},
    issn = {0004-6264},
    doi = {10.1093/pasj/39.3.457},
    url = {https://doi.org/10.1093/pasj/39.3.457},
}

@article{Perez:1996ti,
    author = "Perez, Christopher A. and Silbergleit, Alexander S. and Wagoner, Robert V. and Lehr, Dana E.",
    title = "{Relativistic diskoseismology. 1. Analytical results for 'gravity modes'}",
    eprint = "astro-ph/9601146",
    archivePrefix = "arXiv",
    reportNumber = "SU-ITP-96-1",
    doi = "10.1086/303658",
    journal = "Astrophys. J.",
    volume = "476",
    pages = "589--604",
    year = "1997"
}

@article{Wagoner:1998hh,
    author = "Wagoner, Robert V.",
    title = "{Relativistic diskoseismology}",
    eprint = "astro-ph/9805028",
    archivePrefix = "arXiv",
    doi = "10.1016/S0370-1573(98)00104-5",
    journal = "Phys. Rept.",
    volume = "311",
    pages = "259",
    year = "1999"
}

@article{Abramowicz:2011xu,
    author = "Abramowicz, Marek A. and Fragile, P. Chris",
    title = "{Foundations of Black Hole Accretion Disk Theory}",
    eprint = "1104.5499",
    archivePrefix = "arXiv",
    primaryClass = "astro-ph.HE",
    reportNumber = "NSF-KITP-12-055",
    doi = "10.12942/lrr-2013-1",
    journal = "Living Rev. Rel.",
    volume = "16",
    pages = "1",
    year = "2013"
}

@article{Brink:2015roa,
    author = "Brink, Jeandrew and Geyer, Marisa and Hinderer, Tanja",
    title = "{Astrophysics of resonant orbits in the Kerr metric}",
    eprint = "1501.07728",
    archivePrefix = "arXiv",
    primaryClass = "gr-qc",
    doi = "10.1103/PhysRevD.91.083001",
    journal = "Phys. Rev. D",
    volume = "91",
    number = "8",
    pages = "083001",
    year = "2015"
}

@ARTICLE{1992ApJ...393..697N,
       author = {{Nowak}, Michael A. and {Wagoner}, Robert V.},
        title = "{Diskoseismology: Probing Accretion Disks. II. G-Modes, Gravitational Radiation Reaction, and Viscosity}",
      journal = {\apj},
         year = 1992,
        month = jul,
       volume = {393},
        pages = {697},
          doi = {10.1086/171538},
       adsurl = {https://ui.adsabs.harvard.edu/abs/1992ApJ...393..697N}
}

@article{Fu:2008iw,
    author = "Fu, Wen and Lai, Dong",
    title = "{Effects of Magnetic Fields on the Diskoseismic Modes of Accreting Black Holes}",
    eprint = "0806.1938",
    archivePrefix = "arXiv",
    primaryClass = "astro-ph",
    doi = "10.1088/0004-637X/690/2/1386",
    journal = "Astrophys. J.",
    volume = "690",
    pages = "1386--1392",
    year = "2009"
}

@article{Reynolds:2008wp,
    author = "Reynolds, Christopher S. and Miller, M. Coleman",
    title = "{On the time variability of geometrically-thin black hole accretion disks I : the search for modes in simulated disks}",
    eprint = "0805.2950",
    archivePrefix = "arXiv",
    primaryClass = "astro-ph",
    doi = "10.1088/0004-637X/692/1/869",
    journal = "Astrophys. J.",
    volume = "692",
    pages = "869--886",
    year = "2009"
}

@article{Ortega-Rodriguez:2015dpa,
    author = "Ortega-Rodr{\'\i}guez, Manuel and Sol{\'\i}s-S{\'a}nchez, Hugo and Arguedas-Leiva, J. Agust{\'\i}n and Wagoner, Robert V. and Levine, Adam",
    title = "{Do Magnetic Fields Destroy Black Hole Accretion Disk g-Modes?}",
    eprint = "1506.08314",
    archivePrefix = "arXiv",
    primaryClass = "astro-ph.HE",
    doi = "10.1088/0004-637X/809/1/15",
    journal = "Astrophys. J.",
    volume = "809",
    number = "1",
    pages = "15",
    year = "2015"
}

@article{Clausen:2012zu,
    author = "Clausen, Drew and Sigurdsson, Steinn and Chernoff, David F.",
    title = "{Black Hole-Neutron Star Mergers in Globular Clusters}",
    eprint = "1210.8153",
    archivePrefix = "arXiv",
    primaryClass = "astro-ph.HE",
    doi = "10.1093/mnras/sts295",
    journal = "Mon. Not. Roy. Astron. Soc.",
    volume = "428",
    pages = "3618",
    year = "2013"
}

@article{Ye:2019xvf,
    author = "Ye, Claire S. and Fong, Wen-fai and Kremer, Kyle and Rodriguez, Carl L. and Chatterjee, Sourav and Fragione, Giacomo and Rasio, Frederic A.",
    title = "{On the Rate of Neutron Star Binary Mergers from Globular Clusters}",
    eprint = "1910.10740",
    archivePrefix = "arXiv",
    primaryClass = "astro-ph.HE",
    doi = "10.3847/2041-8213/ab5dc5",
    journal = "Astrophys. J. Lett.",
    volume = "888",
    number = "1",
    pages = "L10",
    year = "2020"
}

@article{OLeary:2005vqo,
    author = "O'Leary, Ryan M. and Rasio, Frederic A. and Fregeau, John M. and Ivanova, Natalia and O'Shaughnessy, Richard W.",
    title = "{Binary mergers and growth of black holes in dense star clusters}",
    eprint = "astro-ph/0508224",
    archivePrefix = "arXiv",
    doi = "10.1086/498446",
    journal = "Astrophys. J.",
    volume = "637",
    pages = "937--951",
    year = "2006"
}

@article{Sedda:2020wzl,
    author = "Sedda, Manuel Arca",
    title = "{Dissecting the properties of neutron star - black hole mergers originating in dense star clusters}",
    eprint = "2003.02279",
    archivePrefix = "arXiv",
    primaryClass = "astro-ph.GA",
    doi = "10.1038/s42005-020-0310-x",
    journal = "Commun. Phys.",
    volume = "3",
    pages = "43",
    year = "2020"
}

@article{Iozzo:2020jcu,
    author = "Iozzo, Dante A. B. and Boyle, Michael and Deppe, Nils and Moxon, Jordan and Scheel, Mark A. and Kidder, Lawrence E. and Pfeiffer, Harald P. and Teukolsky, Saul A.",
    title = "{Extending gravitational wave extraction using Weyl characteristic fields}",
    eprint = "2010.15200",
    archivePrefix = "arXiv",
    primaryClass = "gr-qc",
    doi = "10.1103/PhysRevD.103.024039",
    journal = "Phys. Rev. D",
    volume = "103",
    number = "2",
    pages = "024039",
    year = "2021"
}

@article{Hempel:2009mc,
    author = "Hempel, Matthias and Schaffner-Bielich, Jurgen",
    title = "{Statistical Model for a Complete Supernova Equation of State}",
    eprint = "0911.4073",
    archivePrefix = "arXiv",
    primaryClass = "nucl-th",
    doi = "10.1016/j.nuclphysa.2010.02.010",
    journal = "Nucl. Phys. A",
    volume = "837",
    pages = "210--254",
    year = "2010"
}

@article{Yang:2025jfx,
    author = {Yang, Xing and L{\"u}, Hou-Jun and Rice, Jared and Liang, En-Wei},
    title = "{Discovery of high-frequency quasi-periodic oscillation in short-duration gamma-ray bursts}",
    eprint = "2501.14207",
    archivePrefix = "arXiv",
    primaryClass = "astro-ph.HE",
    doi = "10.1093/mnras/staf162",
    journal = "Mon. Not. Roy. Astron. Soc.",
    volume = "537",
    number = "3",
    pages = "2313--2322",
    year = "2025"
}

@article{Chirenti:2023dzl,
    author = "Chirenti, Cecilia and Dichiara, Simone and Lien, Amy and Miller, M. Coleman and Preece, Robert",
    title = "{Kilohertz quasiperiodic oscillations in short gamma-ray bursts}",
    eprint = "2301.02864",
    archivePrefix = "arXiv",
    primaryClass = "astro-ph.HE",
    doi = "10.1038/s41586-022-05497-0",
    journal = "Nature",
    volume = "613",
    number = "7943",
    pages = "253--256",
    year = "2023"
}

@article{Siegel:2022gwc,
    author = "Siegel, Jared C. and others",
    title = "{Investigating the Lower Mass Gap with Low-mass X-Ray Binary Population Synthesis}",
    eprint = "2209.06844",
    archivePrefix = "arXiv",
    primaryClass = "astro-ph.HE",
    doi = "10.3847/1538-4357/ace9d9",
    journal = "Astrophys. J.",
    volume = "954",
    number = "2",
    pages = "212",
    year = "2023"
}

@article{Hinderer:2018pei,
    author = "Hinderer, Tanja and others",
    title = "{Distinguishing the nature of comparable-mass neutron star binary systems with multimessenger observations: GW170817 case study}",
    eprint = "1808.03836",
    archivePrefix = "arXiv",
    primaryClass = "astro-ph.HE",
    doi = "10.1103/PhysRevD.100.063021",
    journal = "Phys. Rev. D",
    volume = "100",
    number = "6",
    pages = "06321",
    year = "2019"
}

@article{Chase:2021ood,
    author = "Chase, Eve A. and O'Connor, Brendan and Fryer, Christopher L. and Troja, Eleonora and Korobkin, Oleg and Wollaeger, Ryan T. and Ristic, Marko and Fontes, Christopher J. and Hungerford, Aimee L. and Herring, Angela M.",
    title = "{Kilonova Detectability with Wide-field Instruments}",
    eprint = "2105.12268",
    archivePrefix = "arXiv",
    primaryClass = "astro-ph.HE",
    reportNumber = "LA-UR-21-24587",
    doi = "10.3847/1538-4357/ac3d25",
    journal = "Astrophys. J.",
    volume = "927",
    number = "2",
    pages = "163",
    year = "2022"
}

@article{Lousto:2009mf,
    author = "Lousto, Carlos O. and Campanelli, Manuela and Zlochower, Yosef and Nakano, Hiroyuki",
    editor = "Husa, Sascha and Krishnan, Badri",
    title = "{Remnant Masses, Spins and Recoils from the Merger of Generic Black-Hole Binaries}",
    eprint = "0904.3541",
    archivePrefix = "arXiv",
    primaryClass = "gr-qc",
    doi = "10.1088/0264-9381/27/11/114006",
    journal = "Class. Quant. Grav.",
    volume = "27",
    pages = "114006",
    year = "2010"
}

@article{Buonanno:2007sv,
    author = "Buonanno, Alessandra and Kidder, Lawrence E. and Lehner, Luis",
    title = "{Estimating the final spin of a binary black hole coalescence}",
    eprint = "0709.3839",
    archivePrefix = "arXiv",
    primaryClass = "astro-ph",
    doi = "10.1103/PhysRevD.77.026004",
    journal = "Phys. Rev. D",
    volume = "77",
    pages = "026004",
    year = "2008"
}

@article{Gupta:2019nwj,
    author = "Gupta, Anuradha and Gerosa, Davide and Arun, K. G. and Berti, Emanuele and Farr, Will M. and Sathyaprakash, B. S.",
    title = "{Black holes in the low mass gap: Implications for gravitational wave observations}",
    eprint = "1909.05804",
    archivePrefix = "arXiv",
    primaryClass = "gr-qc",
    reportNumber = "LIGO-P1900271",
    doi = "10.1103/PhysRevD.101.103036",
    journal = "Phys. Rev. D",
    volume = "101",
    number = "10",
    pages = "103036",
    year = "2020"
}

@article{Samsing:2020qqd,
    author = "Samsing, Johan and Hotokezaka, Kenta",
    title = "{Populating the Black Hole Mass Gaps in Stellar Clusters: General Relations and Upper Limits}",
    eprint = "2006.09744",
    archivePrefix = "arXiv",
    primaryClass = "astro-ph.HE",
    doi = "10.3847/1538-4357/ac2b27",
    journal = "Astrophys. J.",
    volume = "923",
    number = "1",
    pages = "126",
    year = "2021"
}

@article{Pfeiffer:2007yz,
    author = "Pfeiffer, Harald P. and Brown, Duncan A. and Kidder, Lawrence E. and Lindblom, Lee and Lovelace, Geoffrey and Scheel, Mark A.",
    editor = "Campanelli, Manuela and Rezzolla, Luciano",
    title = "{Reducing orbital eccentricity in binary black hole simulations}",
    eprint = "gr-qc/0702106",
    archivePrefix = "arXiv",
    doi = "10.1088/0264-9381/24/12/S06",
    journal = "Class. Quant. Grav.",
    volume = "24",
    pages = "S59--S82",
    year = "2007"
}

@article{Foucart:2015vpa,
    author = "Foucart, Francois and O'Connor, Evan and Roberts, Luke and Duez, Matthew D. and Haas, Roland and Kidder, Lawrence E. and Ott, Christian D. and Pfeiffer, Harald P. and Scheel, Mark A. and Szilagyi, Bela",
    title = "{Post-merger evolution of a neutron star-black hole binary with neutrino transport}",
    eprint = "1502.04146",
    archivePrefix = "arXiv",
    primaryClass = "astro-ph.HE",
    doi = "10.1103/PhysRevD.91.124021",
    journal = "Phys. Rev. D",
    volume = "91",
    number = "12",
    pages = "124021",
    year = "2015"
}

@article{Etienne:2011ea,
    author = "Etienne, Zachariah B. and Liu, Yuk Tung and Paschalidis, Vasileios and Shapiro, Stuart L.",
    title = "{General relativistic simulations of black hole-neutron star mergers: Effects of magnetic fields}",
    eprint = "1112.0568",
    archivePrefix = "arXiv",
    primaryClass = "astro-ph.HE",
    doi = "10.1103/PhysRevD.85.064029",
    journal = "Phys. Rev. D",
    volume = "85",
    pages = "064029",
    year = "2012"
}

@article{Gonzalez:2022mgo,
    author = "Gonzalez, Alejandra and others",
    title = "{Second release of the CoRe database of binary neutron star merger waveforms}",
    eprint = "2210.16366",
    archivePrefix = "arXiv",
    primaryClass = "gr-qc",
    doi = "10.1088/1361-6382/acc231",
    journal = "Class. Quant. Grav.",
    volume = "40",
    number = "8",
    pages = "085011",
    year = "2023"
}

@dataset{markin_2026_18374717,
  author       = {Markin, Ivan and
                  Dietrich, Tim},
  title        = {Numerical simulations of black hole-neutron star
                   mergers with equal and near-equal mass ratios -
                   Initial data
                  },
  month        = jan,
  year         = 2026,
  publisher    = {Zenodo},
  doi          = {10.5281/zenodo.18374717},
  url          = {https://doi.org/10.5281/zenodo.18374717},
}

@dataset{markin_2026_18377211,
  author       = {Markin, Ivan and
                  Dietrich, Tim},
  title        = {Numerical simulations of black hole-neutron star
                   mergers with equal and near-equal mass ratios --
                   Waveforms
                  },
  month        = jan,
  year         = 2026,
  publisher    = {Zenodo},
  doi          = {10.5281/zenodo.18377211},
  url          = {https://doi.org/10.5281/zenodo.18377211},
}

@article{Typel:2013rza,
    author = {Typel, S. and Oertel, M. and Kl{\"a}hn, T.},
    title = "{CompOSE CompStar online supernova equations of state harmonising the concert of nuclear physics and astrophysics compose.obspm.fr}",
    eprint = "1307.5715",
    archivePrefix = "arXiv",
    primaryClass = "astro-ph.SR",
    doi = "10.1134/S1063779615040061",
    journal = "Phys. Part. Nucl.",
    volume = "46",
    number = "4",
    pages = "633--664",
    year = "2015"
}

@article{Oertel:2016bki,
    author = {Oertel, M. and Hempel, M. and Kl{\"a}hn, T. and Typel, S.},
    title = "{Equations of state for supernovae and compact stars}",
    eprint = "1610.03361",
    archivePrefix = "arXiv",
    primaryClass = "astro-ph.HE",
    doi = "10.1103/RevModPhys.89.015007",
    journal = "Rev. Mod. Phys.",
    volume = "89",
    number = "1",
    pages = "015007",
    year = "2017"
}

@article{CompOSECoreTeam:2022ddl,
    author = "Typel, S. and others",
    collaboration = "CompOSE Core Team",
    title = "{CompOSE Reference Manual}",
    eprint = "2203.03209",
    archivePrefix = "arXiv",
    primaryClass = "astro-ph.HE",
    doi = "10.1140/epja/s10050-022-00847-y",
    journal = "Eur. Phys. J. A",
    volume = "58",
    number = "11",
    pages = "221",
    year = "2022"
}

@article{Bauswein:2010dn,
    author = "Bauswein, A. and Janka, H. -Th. and Oechslin, R.",
    title = "{Testing Approximations of Thermal Effects in Neutron Star Merger Simulations}",
    eprint = "1006.3315",
    archivePrefix = "arXiv",
    primaryClass = "astro-ph.SR",
    doi = "10.1103/PhysRevD.82.084043",
    journal = "Phys. Rev. D",
    volume = "82",
    pages = "084043",
    year = "2010"
}

@article{Romani:2022jhd,
    author = "Romani, Roger W. and Kandel, D. and Filippenko, Alexei V. and Brink, Thomas G. and Zheng, WeiKang",
    title = "{PSR J0952{\ensuremath{-}}0607: The Fastest and Heaviest Known Galactic Neutron Star}",
    eprint = "2207.05124",
    archivePrefix = "arXiv",
    primaryClass = "astro-ph.HE",
    doi = "10.3847/2041-8213/ac8007",
    journal = "Astrophys. J. Lett.",
    volume = "934",
    number = "2",
    pages = "L17",
    year = "2022"
}

@article{Kiziltan:2013oja,
    author = {Kiziltan, B{\"u}lent and Kottas, Athanasios and De Yoreo, Maria and Thorsett, Stephen E.},
    title = "{The Neutron Star Mass Distribution}",
    eprint = "1309.6635",
    archivePrefix = "arXiv",
    primaryClass = "astro-ph.SR",
    doi = "10.1088/0004-637X/778/1/66",
    journal = "Astrophys. J.",
    volume = "778",
    pages = "66",
    year = "2013"
}

@article{OConnor:2009iuz,
    author = "O'Connor, Evan and Ott, Christian D.",
    editor = "Ott, Christian D. and Pethick, C. J. and Rezzolla, Luciano",
    title = "{A New Open-Source Code for Spherically-Symmetric Stellar Collapse to Neutron Stars and Black Holes}",
    eprint = "0912.2393",
    archivePrefix = "arXiv",
    primaryClass = "astro-ph.HE",
    doi = "10.1088/0264-9381/27/11/114103",
    journal = "Class. Quant. Grav.",
    volume = "27",
    pages = "114103",
    year = "2010"
}

@article{Bardeen:1972fi,
    author = "Bardeen, James M. and Press, William H. and Teukolsky, Saul A",
    title = "{Rotating black holes: Locally nonrotating frames, energy extraction, and scalar synchrotron radiation}",
    reportNumber = "OAP-288",
    doi = "10.1086/151796",
    journal = "Astrophys. J.",
    volume = "178",
    pages = "347",
    year = "1972"
}

\end{document}